\documentclass[journal]{IEEEtran}
\usepackage{morewrites}
\usepackage[T1]{fontenc}
\usepackage[utf8]{inputenc}
\usepackage{microtype}
\usepackage[english]{babel}
\usepackage{xcolor}
\usepackage[fleqn]{amsmath}
\usepackage{amsfonts, amssymb, yhmath}
\usepackage{amsthm}
\usepackage{thmtools}
\usepackage{bm}
\usepackage{subcaption}
\usepackage{tikz}
\usepackage{pgfplots}
\pgfplotsset{compat=1.16}
\usepackage{hyperref}
\hypersetup{colorlinks=true,allcolors=black}
\DeclareMathOperator{\rect}{rect}
\usepackage{algorithm}
\usepackage{algorithmicx}
\usepackage{algpseudocode}

\usepackage[nolist]{acronym}
\usepackage[nameinlink,capitalize]{cleveref}

\usepackage{csquotes}
\usepackage{siunitx}
\DeclareUnicodeCharacter{2212}{−}
\usepgfplotslibrary{groupplots,dateplot}
\usetikzlibrary{patterns,shapes.arrows}
\pgfplotsset{compat=newest}

\usepackage{booktabs}

\usepackage[fleqn]{amsmath}
\usepackage{amsfonts, amssymb, yhmath}
\usepackage{amsthm}
\usepackage{thmtools}
\usepackage{bm}
\usepackage{graphicx}
\usepackage[]{pgfplots}
\pgfplotsset{compat=newest}
\usepackage[]{glossaries}
\usepackage[]{hyperref}
\usepackage[]{cleveref}

\begin{acronym}

\acro{ISAC}{Integrated Sensing and Communications}%
\acro{JCAS}{Joint Communication and Sensing}
\acro{V2X}{vehicle-to-everything}
\acro{SPORT}{SParse Offset-based Regular InTerleaving}
\acro{SRADIO}{Sparse RAndom Dynamic Interleaving without Offset}
\acro{B5G}{beyond fifth-generation}
\acro{6G}{sixth-generation}
\acro{OFDM}{orthogonal frequency division multiplexing}
\acro{OFDMA}{orthogonal frequency division multiple access}
\acro{ITS}{Intelligent Transportation System}
\acro{C-V2X}{cellular-vehicle-to-everything}
\acro{NOMP}{Newtonized orthogonal matching pursuit}
\acro{2D-FFT}{two dimensional fast Fourier transform}
\acro{CP}{cyclic prefix}
\acro{CS}{compressed sensing}
\acro{OMP}{orthogonal matching pursuit}
\acro{DFT}{Discrete Fourier Transform}
\acro{BP}{Basis Pursuit}
\acro{SALSA}{split augmented Lagrangian shrinkage}
\acro{AMP}{approximate message passing}
\acro{ANM}{atomic norm minimization}
\acro{SWPR}{strong-to-weak-target peak ratio}
\acro{AWGN}{additive white Gaussian noise}
\acro{RMSE}{root-mean-squared-error}
\acro{RCS}{radar cross-section}
\acro{VISTA}{Virtual Road Simulation and Test Area}
\acro{SIMFE}{sparse iterative multidimensional frequency estimation}
\acro{FFT}{fast Fourier transform}
\acro{MUSIC}{multiple signal classification}
\acro{ESPRIT}{estimation of signal parameters via rational invariance techniques}
\acro{PSLR}{peak-to-sidelobe ratio}
\acro{MC}{Monte Carlo}
\acro{MCRB}{modified Cramer-Rao bound}
\acro{CRB}{Cramer-Rao bound}
\acro{LoS}{line-of-sight}
\end{acronym}
\title{Newtonized Orthogonal Matching Pursuit for High-Resolution Target Detection in Sparse OFDM ISAC Systems}%
\author{%
  Syed Najaf Haider Shah, Sebastian~Semper, Aamir Ullah Khan, Christian Schneider, and Joerg Robert
}%
\IEEEaftertitletext{This work has been submitted to the IEEE for possible publication. Copyright may be transferred without notice, after which this version may no longer be accessible.}
\begin{document}

\maketitle
\IEEEpeerreviewmaketitle
\begin{abstract}
Integrated Sensing and Communication (ISAC) is a technology paradigm that combines sensing capabilities with communication functionalities in a single device or system.
In vehicle-to-everything (V2X) sidelink, ISAC can provide enhanced safety by allowing vehicles to not only communicate with one another but also sense the surrounding environment by using sidelink signals.
In ISAC-capable V2X sidelink, the random resource allocation results in an unstructured and sparse distribution of time and frequency resources in the received orthogonal frequency division multiplexing (OFDM) grid, leading to degraded radar detection performance when processed using the conventional 2D-FFT method.
To address this challenge, this paper proposes a high-resolution off-grid radar target detection algorithm irrespective of the OFDM grid structure.
The proposed method utilizes the Newtonized orthogonal matching pursuit (NOMP) algorithm to effectively detect weak targets masked by the sidelobes of stronger ones and accurately estimates off-grid range and velocity parameters with minimal resources through Newton refinements.
Simulation results demonstrate the superior performance of the proposed NOMP-based target detection algorithm compared to existing compressed sensing (CS) methods in terms of detection probability, resolution, and accuracy.
Additionally, experimental validation is performed using a bi-static radar setup in a semi-anechoic chamber.
The measurement results validate the simulation findings, showing that the proposed algorithm significantly enhances target detection and parameter estimation accuracy in realistic scenarios.
\end{abstract}
\section{Introduction}\label{introduction}
\IEEEPARstart{I}{n} recent decades, the fields of wireless communications and radar sensing have witnessed significant advancements, driven by the growing demand for high-speed data transfer and precise environmental awareness.
However, the rapid increase in connected devices and applications, such as autonomous driving, smart cities, and extended reality, has put immense pressure on the already crowded electromagnetic spectrum.
This spectrum scarcity has led to the emergence of \ac{ISAC} as a promising solution that merges communication and radar sensing functionalities within a single system.
\ac{ISAC} enables the efficient sharing of limited spectrum and hardware resources between these two domains, resulting in substantial cost savings and improved system performance~\cite{Thoma2021JointOverview, Du2023AnCommunication}.

\ac{ISAC} is not just about spectrum efficiency, it is also a key enabler for next-generation cellular technologies like \ac{B5G} and \ac{6G}.
\ac{ISAC} represents a promising key technology for enhancing connectivity and intelligent decision-making in various applications such as high-resolution sensing and tracking, simultaneous localization, mapping, and imaging~\cite{PinTan2021IntegratedDirections} etc.
In \ac{C-V2X}, particularly in 6G-V2X systems, \ac{ISAC} will enable vehicles to communicate with one another while simultaneously sensing their environment.
This dual capability is crucial for facilitating cooperative driving, where vehicles can share information about their surroundings, road conditions, and potential hazards in real-time~\cite{Cheng2022IntegratedVCN}.
By integrating sensing and communication functions, \ac{ISAC} enhances situational awareness, enabling vehicles to make informed decisions that improve overall traffic safety ultimately contributing to safer, more efficient, and intelligent transportation systems.

In a multiuser \ac{ISAC}-capable \ac{V2X} sidelink system, vehicles in autonomous mode randomly select distinct frequency (subcarriers) and time (symbols) resources from an \ac{OFDM} resource grid to perform communication and radar sensing simultaneously~\cite{Shah2023Radar-EnabledCommunication}.
This random resource selection is susceptible to mutual interference and resource collisions due to the hidden-node problem~\cite{Todisco2021PerformanceSimulator}.
When the interfered or overlapped resources are removed from this unstructured \ac{OFDM} resource grid, the received \ac{OFDM} signal at the radar receiver becomes sparse in both time and frequency domain.
This sparsity poses significant challenges for conventional radar processing methods, such as \ac{2D-FFT}, \ac{MUSIC}, or \ac{ESPRIT}, which are typically designed for either continuous or regularly spaced time-frequency \ac{OFDM} signals~\cite{Rahman2019JointSensing}.
Applying these algorithms to sparse and unstructured \ac{OFDM} signals results in degraded target detection performance, where unpredictable range-Doppler sidelobes appear in the radar map. These sidelobes obscure weaker targets, making target detection difficult and unreliable~\cite{Thoma2019CooperativeSafety}.

To address the challenges posed by the sparse nature of the \ac{OFDM} signal in \ac{ISAC} systems, \ac{CS} techniques offer a promising alternative.
\ac{CS} has been extensively utilized in both communication and radar sensing applications.
In communication systems, \ac{CS} is employed for channel estimation through sparse signal recovery~\cite{Wan2020CompressiveLens-Array, Choi2017CompressedTricks}, while in radar sensing, it provides super-resolution and enhanced accuracy in parameter estimation~\cite{Hadi2015CompressiveOverview,Ender2010OnRadar}.
Notably, \ac{CS} methods have demonstrated superiority over subspace-based approaches like \ac{MUSIC} in noisy environments, particularly when strong dominant clutter or direct signals are present~\cite{Berger2010SignalWaveforms}.
Additionally, \ac{CS} is well-suited for scenarios where continuous target observation is not feasible, as it can reconstruct signals from a small number of samples by exploiting signal sparsity, even when the sampling interval is irregular.
However, \ac{CS} methods are not without challenges, particularly the grid mismatch problem.
This issue arises when the true parameters of a target (e.g., range and Doppler shift) do not align with the discretized grid used for parameter estimation~\cite{Bhaskar2013AtomicEstimation}.
Even with a finely discretized parameter grid, the true target parameters often fall between grid points, leading to estimation errors~\cite{Chi2011SensitivitySensing}.
Furthermore, increasing the grid resolution to mitigate this issue results in a quadratic increase in computational complexity, making real-time processing impractical.

Addressing these challenges is essential to unlocking the full potential of \ac{ISAC} in \ac{V2X} systems and other advanced applications.
Effective radar target detection in dynamic environments requires low-complexity algorithms that offer super-resolution capabilities, eliminate grid mismatches, and accurately estimate both on-grid and off-grid parameters.
The development of such algorithms will be key to enabling reliable, high-precision sensing in \ac{ISAC}-capable \ac{V2X} systems, paving the way for safer, smarter, and more connected transportation environments.
\subsection{Related Work}\label{sota}

Several approaches have been proposed to address the challenges posed by sparse \ac{OFDM} signals in \ac{ISAC} systems.
The authors in~\cite{Hakobyan2016ASensing,Hakobyan2016AInterleaving} introduced a \ac{CS} method named \ac{SIMFE} based on the \ac{OMP} algorithm to deal with the increased range sidelobes caused by non-equidistant subcarrier interleaving in the \ac{OFDM} grid.
This method exploits the idea of interpolation for an off-grid frequency estimation.
However, this method focuses solely on range sidelobe suppression and does not account for sparsity in the time domain.
Other approaches, such as those in~\cite{Nuss2017ASensing,Kong2017SparseNetworks}, combine the \ac{DFT} and \ac{BP} algorithms to resolve ambiguities in range estimation due to random subcarrier allocation but suffer from high computational complexity.
To enhance range and velocity profile reconstruction with sparse OFDM signals, \ac{AMP} and \ac{SALSA} algorithms were employed in~\cite{Knill2018HighRadar,Knill2019RandomReconstruction}.
However, these algorithms struggle when sparsity across \ac{OFDM} symbols is considered.
Moreover, these methods do not address the grid mismatch problem, which arises when true parameter values do not align with the discretized grid.
To account for this issue, gridless \ac{CS} methods based on \ac{ANM} are employed to estimate off-grid parameters accurately~\cite{Tang2013CompressedGrid,Semper2019ADMMEstimation}, but their computational complexity is often prohibitive for practical implementation.

While existing \ac{CS} algorithms for sparse OFDM signals have made important strides, they exhibit several key shortcomings.
Some suffer from high Doppler sidelobes due to the sparsity along the time domain not being considered.
Others fail to address the grid mismatch problem.
Additionally, those that offer solutions to grid mismatch issues suffer from high computational complexity, limiting their practical use in real-time applications.
\subsection{Contributions}\label{contribution}

To address these shortcomings, this paper proposes a super-resolution off-grid radar target detection algorithm based on the \ac{NOMP} regardless of the \ac{OFDM} grid structure.
The proposed algorithm effectively detects weak targets masked by the sidelobes of strong targets in the sparse \ac{OFDM}-based multiuser \ac{ISAC}-capable \ac{V2X} sidelink system with minimal sensing resources.

The key contributions of this paper are summarized as follows:
\begin{enumerate}
    \item We introduce a high-resolution radar target detection algorithm that operates without restrictions on the structure of the OFDM grid, leveraging the \ac{NOMP} method for both on-grid and off-grid parameter estimation.
    The algorithm consists of three key steps: (1) coarse on-grid estimation, (2) local Newton refinements, and (3) joint Newton refinements.
    In the first step, standard \ac{OMP} is employed to estimate the on-grid delay and Doppler shift parameters.
    To accelerate the process, the conventional correlation step in the \ac{OMP} is replaced with a computationally efficient \ac{2D-FFT}, ensuring faster on-grid estimation.
    In the second step, Newton refinements are applied iteratively to the newly detected on-grid parameter estimation, allowing for accurate off-grid parameter estimation.
    In the third step, joint Newton refinements are applied to all detected parameters to obtain further fine-tuned estimates.
    This feedback mechanism enables the re-evaluation of previously estimated parameters based on newly detected ones, improving estimation accuracy and handling target interference.
    To reduce the computational burden associated with joint Newton refinements, the block Hessian matrix is simplified into a block diagonal matrix, yielding a low-complexity solution.
    These enhancements make the proposed algorithm not only highly accurate but also computationally efficient, rendering it suitable for real-time radar target detection in multiuser \ac{ISAC} systems.
    Moreover, the proposed algorithm does not require a priori knowledge of the number of targets for termination rather it supports a CFAR-based stopping criterion as given in~\cite{Mamandipoor2016NewtonizedContinuum}.
    \item The proposed algorithm is rigorously evaluated through simulations to assess its performance across several key metrics, including effective weak target recovery, detection of closely-spaced off-grid targets, \ac{RMSE} in range and velocity estimations, convergence behavior, and execution time.
    The results demonstrate the superiority of the proposed \ac{NOMP}-based algorithm compared to conventional methods such as \ac{2D-FFT}, \ac{OMP}, and \ac{SALSA}, highlighting its enhanced ability to resolve targets under challenging conditions.
    \item To evaluate the performance of the proposed algorithm in realistic conditions, we conducted experiments using a bi-static radar setup within a semi-anechoic chamber.
    The experiments were facilitated by the \ac{VISTA} at the Thuringian Center of Innovation in Mobility, providing a controlled environment for precise evaluation~\cite{Dobereiner2019JointMeasurements}.
    The measurement results corroborate the simulation findings, demonstrating the practical feasibility of the \ac{NOMP}-based algorithm in real-world scenarios.
\end{enumerate}
\subsection{Notation}\label{notation}

Scalars, vectors, matrices, and sets are denoted using different typographical conventions.
Scalars are represented by italic letters ($a$) or Greek letters ($\alpha$).
Vectors are indicated by bold lowercase letters ($\mathbf{a}$), and matrices are denoted by bold uppercase letters ($\mathbf{A}$).
Matrix operations follow standard mathematical conventions.
$\mathbf{A}^{\mathrm{T}}$, $\mathbf{A}^{\mathrm{H}}$, $\mathbf{A}^{-1}$,$\mathbf{A}^{\dagger}$, and $\mathbf{A}^{\ast}$ represent transpose, Hermitian transpose, inverse, pseudo-inverse, and conjugate of matrix $\mathbf{A}$ respectively.
The identity matrix of size $N \times N$ is denoted by $\mathbf{I}_N$.
For element-wise operations, the Hadamard (element-wise) product between matrices $\mathbf{A}$ and $\mathbf{B}$ is represented by $\mathbf{A} \circ \mathbf{B}$, and element-wise division is denoted as $\mathbf{A} \oslash \mathbf{B}$.
The Kronecker product of two matrices $\mathbf{A}$ and $\mathbf{B}$ is denoted as $\mathbf{A} \otimes \mathbf{B}$.
The Fourier transform operation is denoted by $\mathcal{F}\{\cdot\}$, and its inverse by $\mathcal{F}^{-1}\{\cdot\}$.
$\mathcal{O}(\cdot)$ denotes the big-O notation.
Norms and other mathematical notations include the $\ell_2$ norm, which is denoted as  $\left\Vert\mathbf{x}\right\Vert_2$ and represents the Euclidean norm of vector $\mathbf{x}$.

The rest of the paper is organized as follows: Section~\ref{model} presents the \ac{ISAC} system model and signal model.
Section~\ref{estimation} describes the proposed high-resolution off-grid target detection algorithm using \ac{NOMP}.
Section~\ref{simulation} presents the simulation results.
Measurement setup and measurement results are presented in Section~\ref{measurement}.
Finally, Section \ref{conclusion} concludes the paper.
\section{ISAC-capable Sidelink System}\label{model}
\begin{figure}[htbp]
    \centering
    \includegraphics[width=3.4in,height=\textwidth,keepaspectratio]{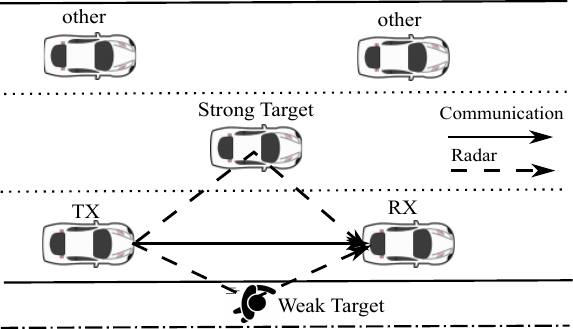}
    \caption{\ac{ISAC}-capable sidelink scenario}
    \label{fig:exampleScene}
\end{figure}

\begin{figure}[htbp]
    \centering
    \includegraphics[width=3.4in,height=\textwidth,keepaspectratio]{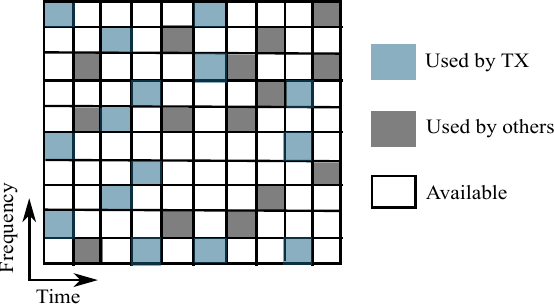}
    \caption{The resource allocation grid obtained by random resource selection in a multiuser \ac{ISAC}-capable sidelink system.}
    \label{fig:sparseofdm}
\end{figure}

We assume a scenario that consists of multiple vehicles moving on a road.
The vehicles exchange data with one another through sidelink signals while also utilizing reflected sidelink signals to sense the surrounding environment, enabling a multiuser \ac{ISAC}-capable sidelink.
As shown in \cref{fig:exampleScene}, in compliance with the half-duplex mode of operation in \ac{V2X} sidelink communication, a transmitting vehicle (TX) transmits data packets to a receiving vehicle (RX) over the sidelink channel.
The RX receives not only the direct sidelink signal from the TX but also the reflected signals from the neighbor vehicle (Strong Target) and a pedestrian (Weak Target).
This configuration enables the RX to operate as a bi-static radar, utilizing the reflected sidelink signals to estimate the range and relative velocity of the targets~\cite{Khan2024TargetModel}.
The other vehicles (other) also communicate with the RX through the \ac{V2X} sidelink multiple access technology.
In this multiuser \ac{ISAC}-capable sidelink setup, the TX randomly selects time and frequency resources from the \ac{OFDM} resource pool which is also shared by other vehicles as shown in \cref{fig:sparseofdm}.
This random selection forms an unstructured sparse \ac{OFDM} grid which will be received by the RX for communication and radar processing.
The TX also sends the information of its time-frequency resources (i.e., number and locations in the sparse \ac{OFDM} grid) to the RX over the sidelink control channel so that RX have the knowledge of the TX signal.

\subsection{Signal Model}

The transmitted signal $x : \mathbb{R} \rightarrow \mathbb{C}$ in complex baseband consists of $N$ subcarriers and $M$ OFDM symbols resulting in an $N\times M$ TX resource grid $\mathbf{X}$. 
Within $\mathbf{X}$, the TX selects a random subset $\Omega_{\rm{s}}$ of resources from $\Omega$ where $\Omega=\{(n,m) \mid 0<n\leq N , 0<m\leq M\}$ and $\Omega_{\rm{s}} \subset \Omega$.
Hence, the transmit signal $x$ is given by
\begin{equation} 
    x(t) = 
        \sum_{m=0}^{M-1}\left ({
            \sum_{n=0}^{N-1} 
                \mathbf{X}_{n,m}{\mathrm{e}}^{\jmath\frac {2\pi}{T}nt}}
        \right)
        \rect_T(t-mT_o),
    \label{tx_signal}
\end{equation}
with
\begin{equation} 
    \mathbf{X}_{n,m} \in 
    \begin{cases} 
        \mathcal{A}_{n, m}, & \text{if}\ \{n,m\}\in \Omega_{\rm{s}}\\ 
        0, & \text{otherwise}
    \end{cases} 
    \label{resource_element} 
\end{equation}
where $\mathcal{A}_{n, m}$, $T$, and $T_o$ denote the modulation symbol (e.g., QPSK), rectangular pulse duration, and the \ac{OFDM} symbol duration including the \ac{CP} respectively. 
The received \ac{OFDM} grid at the RX after removing the \ac{CP} and taking the \ac{DFT} can be written as
\begin{equation}
    \mathbf{Y}_{n,m} = \sum\limits_{k = 1}^{{K}} {{\beta_{k}}{\mathbf{X}_{n,m}}{{\mathrm{e}}^{-\jmath 2\pi n\Delta f{\tau_k}}}{{\mathrm{e}}^{\jmath 2\pi m T_o \alpha_k}} + {\mathbf{Z}_{n,m}}},
    \label{rx_signal}
\end{equation}
where $K$ is the total number of targets which are assumed to be placed at arbitrary locations in the parameter space, and $\tau_k$ and $\alpha_k$ are bi-static delay and Doppler shift of the $k^{\text{th}}$ target at the RX respectively. 
The term $\beta_k=\vert\beta_k\vert\mathrm{e}^{j\phi_k}$ is the complex attenuation coefficient which includes the attenuation constant and the phase shift of the signal reflected from the $k^{\text{th}}$ target, and ${\mathbf{Z}_{n,m}}$ represents \ac{AWGN} with zero mean.
\section{Off-grid Target Detection using NOMP Algorithm}\label{estimation}
This section introduces the proposed \ac{NOMP}-based target detection algorithm that estimates the targets placed across a continuous parameter space as an off-grid sparse recovery method. 
First, we account for the transmitted symbols from the received \ac{OFDM} signal $\mathbf{Y}$ by element-wise division with $\mathbf{X}$ as given in~\cite{Sturm2011WaveformSensing,Ravelo2023SensingCommunications}, to obtain the frequency domain channel transfer function matrix ${\mathbf{H}}\in \mathbb{C}^{M\times N}$ as
\begin{equation}
    \begin{aligned}
        \mathbf{H}_{n,m}
            &=\frac{\mathbf{Y}_{n,m}}{\mathbf{X}_{n,m}}\\
            &= \sum\limits_{k = 1}^{{K}} 
                {{\beta_{k}}
                {\mathrm{e}^{-\jmath 2\pi n\Delta f{\tau_k}}}
                {\mathrm{e}^{\jmath 2\pi m T_o \alpha_k}}
            + {\mathbf{\Tilde{Z}}_{n,m}}},
    \end{aligned}
    \label{channel_matrix}
\end{equation}
where $\mathbf{H}$ is a sparse matrix with nonzero entries located at the set of indices $\Omega_{\rm{s}}$. 
By excluding the unused resources (i.e., zero entries), the matrix ${\mathbf{H}}$ is compressed to ${\mathbf{H}_s}$. 
We express the compressed matrix ${\mathbf{H}_s}$ in vector form as
\begin{equation}
    \mathbf{h}_s = \sum\limits_{k = 1}^{{K}}
        {{\beta_{k}}
        \mathbf{a}^{\Omega_{\rm{s}}}(\tau_k, \alpha_k)}
        + \mathbf{\Tilde{z}},
     \label{channel_vector}
\end{equation}
where $\mathbf{a}^{\Omega_{\rm{s}}}(\tau_k, \alpha_k)\in \mathbb{C}^{|\Omega_{\rm{s}}| }$ is the response vector corresponding to the $k^{\text{th}}$ off-grid target, i.e.,
\begin{equation}
    \mathbf{a}^{\Omega_{\rm{s}}}(\tau_k, \alpha_k) = \left[{\mathrm{e}^{-\jmath 2\pi n\Delta f{\tau_k}}}
    {\mathrm{e}^{\jmath 2\pi m T_o \alpha_k}}\right]_{\{n,m\}\in \Omega_{\rm{s}}}.
\end{equation}
Thus, \eqref{channel_vector} can be rewritten as
\begin{equation}
    \mathbf{h}_s = \mathbf{A}\mathbf{b}
    + \mathbf{\Tilde{z}},
    \label{channel_vector_rewritten}
\end{equation}
where $\mathbf{A} = \left[\mathbf{a}^{\Omega_{\rm{s}}}(\tau_1, \alpha_1), \mathbf{a}^{\Omega_{\rm{s}}}(\tau_2, \alpha_2), \cdots,\mathbf{a}^{\Omega_{\rm{s}}}(\tau_K, \alpha_K)\right]\in \mathbb{C}^{|\Omega_{\rm{s}}|\times K}$, $\mathbf{b} = \left[\beta_1,\beta_2,\cdots,\beta_K\right]\in \mathbb{C}^{K }$, and $\mathbf{\Tilde{z}}\in\mathbb{C}^{|\Omega_{\rm{s}}|}$ is the noise vector.
\subsection{OMP}
Conventional grid-based \ac{CS} methods discretize the parameter plane into $N \times M$ grid points and assume that each grid point corresponds to a potential target's parameter. 
Let $\bar\tau \overset{\Delta}{=} \{\bar\tau_p \in \left[0, \frac{d_{max}}{c}\right] \mid p=1,\cdots,N\}$, and $\bar\alpha \overset{\Delta}{=} \{\bar\alpha_q \in \left[-\frac{2v_{max}}{\lambda},\frac{2v_{max}}{\lambda}\right] \mid q=1,\cdots,M\}$, be the discretized delay and Doppler shift grids respectively, where $d_{max}$ is the sum of the maximum transmitter-to-target range and the maximum target-to-receiver range, $v_{max}$ is the maximum allowed relative velocity, $\lambda$ and $c$ are the wavelength and speed of light respectively. 
Then, the measurement vector $\mathbf{h}_s$ can be expressed as
\begin{equation}
    \mathbf{h}_s = \mathbf{\bar{A}}\mathbf{\bar{b}}
    + \mathbf{\bar{z}},
    \label{channel_grid}
\end{equation}
where $\mathbf{\bar{A}} = \left[\mathbf{a}(\bar\tau_1, \bar\alpha_1), \mathbf{a}(\bar\tau_2, \bar\alpha_1), \cdots,\mathbf{a}(\bar\tau_N, \bar\alpha_M)\right]\in \mathbb{C}^{|\Omega_{\rm{s}}|\times NM}$ is the sensing matrix, $\mathbf{\bar{b}} = \left[\bar\beta_1,\bar\beta_2,\cdots,\bar\beta_{NM}\right]\in \mathbb{C}^{NM }$ is the attenuation coefficient vector of grid points.
In \eqref{channel_grid}, $\mathbf{h}_s$ and $\mathbf{\bar{A}}$ are known while $\mathbf{\bar{b}}$ is unknown.
Since $|\Omega_{\rm{s}}|\ll NM$, \eqref{channel_grid} is underdetermined, hence there is no unique solution without additional assumptions about the considered problem.
When the targets are sparsely occupying the parameter space, that is, non-zero elements in the solution vector $\mathbf{\bar{b}}$ are very few, \eqref{channel_grid} can be solved by utilizing sparse recovery methods, which impose sparsity by appropriate regularization of the obtained solutions.
One of the most prominent \ac{CS} methods is the \ac{OMP} algorithm, where the main idea is to obtain a new support element of $\mathbf{\bar{b}}$ in each iteration, by determining the maximum correlation between the residual and columns of the measurement matrix.
The \ac{OMP} is stated in \cref{omp_algo}.
\begin{algorithm}
\caption{The \ac{OMP} Algorithm}\label{omp_algo}
\begin{algorithmic}[1]
\State \textbf{Input:} $\mathbf{h}_s$, $\mathbf{\bar{A}}$, $K$
\State \textbf{Initial state:} $\mathbf{r}_0 = \mathbf{h}_s$, $\Pi_0 = \emptyset$, $\ell = 0$
\Repeat
    \State $\ell = \ell + 1$
    \State \textbf{Match step:} \Comment{Compute the correlation}
    \State \quad $\mathbf{c}_{\ell} = \mathbf{\bar{A}}^{\mathrm{H}} \mathbf{r}_{\ell-1}$
    \State \textbf{Identify step:} \Comment{Potential basis mismatch here}
    \State \quad $\Pi_{\ell} = \Pi_{\ell-1} \cup \left\{ \arg\max_j |\mathbf{c}_{\ell}(j)|^2 \right\}$
    \State \textbf{Update step:}\Comment{Basis mismatch affects this step}
    \State \quad $\mathbf{\bar{b}}_{\ell} = \arg\min_{\mathbf{v}:\text{supp}(\mathbf{v}) \subseteq \Pi_{\ell}} \Vert\mathbf{h}_s - \mathbf{\bar{A}}\mathbf{v}\Vert_2^2$
    \State \quad $\mathbf{r}_{\ell} = \mathbf{h}_s - \mathbf{\bar{A}}\mathbf{\bar{b}}_{\ell}$\Comment{Update the residue}
\Until{$\ell = K$}
\State \textbf{Output:} $\mathbf{\bar{b}}_K$, $\Pi_K$
\end{algorithmic}
\end{algorithm}

However, the \ac{OMP} suffers from the basis mismatch issue; that is, when the target lies off the grid points constituting $\mathbf{\bar{A}}$ and $\mathbf{\bar{b}}$, its performance degrades. 
In \cref{omp_algo}, the \textit{identify step} selects the atom with the highest correlation to the current residual. 
However, when the target lies between two grid points, the selected atom will not perfectly align with the actual target, leading to a mismatch that is later accounted for by the estimation of spurious ghost paths, i.e., the algorithm proposes detected target locations that have no physical justification.
This mismatch also further propagates to the \textit{update step}, where the residual is updated based on a suboptimal atom selection, thereby reducing the accuracy of the reconstructed signal and also the reliability of the estimated target parameters.
\subsection{NOMP}
To address this limitation, we introduce a high-resolution off-grid target detection method using the \ac{NOMP} algorithm. 
The \ac{NOMP} extends the \ac{OMP} by introducing Newton refinements to mitigate the basis mismatch.
The proposed algorithm roughly involves three steps: 1) making a coarse estimation of a new target's parameters on a discrete grid using the \ac{OMP}, 2) performing local optimization using Newton refinements in the vicinity of the coarse estimate, essentially introducing off-grid estimates, 3) conducting global optimization using Newton refinements for all previously identified targets jointly. 
By introducing these steps, the algorithm mitigates the impact of basis mismatch, resulting in improved performance compared to standard \ac{OMP}, especially when the target lies off the discretized grid points.

Let $\mathcal{C}=\{\tau_{\ell},\alpha_{\ell},\beta_{\ell} \mid \ell=1,\cdots,K\}$ represent the set of parameters to be estimated.
The corresponding residual will be calculated by
\begin{equation}
    {{\mathbf{h}}_r}(\mathcal{C}) = {\mathbf{h}_s} - \sum\limits_{{\ell} = 1}^{K}
    {\beta_{\ell}}\mathbf{a}^{\Omega_{\rm{s}}} \left( {{\tau_{\ell}},{\alpha_{\ell}}} \right).
    \label{residual}
\end{equation}
We obtain the maximum likelihood estimate of $\mathcal{C}$ by minimizing the residual energy $\Vert{{\mathbf{h}}_r}(\mathcal{C})\Vert_2^2$. 
However, it is computationally inefficient to directly minimize the residual w.r.t all parameters. 
Therefore, we estimate the parameters of multiple targets by performing Newtonian gradient descent for each target's parameters individually using the coarse estimation result of \ac{OMP} as on-grid initial estimates.  
\subsubsection{Initial Coarse Estimation}\label{coarse_estimate}
We first perform a coarse detection of $\tau_{\ell}$ and $\alpha_{\ell}$ within respective discrete grids $\bar\tau$ and $\bar\alpha$.
The coarse estimate of $(\tau_{\ell},\alpha_{\ell})$ is obtained by the \ac{OMP} method summarized in \cref{omp_algo}, where the \textit{match step} involves computing the correlations between the residual and the columns of the sensing matrix $\mathbf{\bar{A}}$.
However, when $\mathbf{\bar{A}}$ is a large matrix, this step incurs a high computational cost due to the large number of matrix-vector multiplications.
To reduce the computational complexity, we propose a modification by exploiting the structure of the sensing matrix.
We know that in our sensing scenario the sensing matrix $\mathbf{\bar{A}}$ can be written as:
\begin{equation}
    \mathbf{\bar{A}} = \mathbf{I_s} (\mathbf{F}_m^\ast \otimes \mathbf{F}_n),
\end{equation}
where $\mathbf{I_s} \in \mathbb{C}^{|\Omega_{\rm{s}}|\times NM}$ is a selection matrix, and  $\mathbf{F}_n \in \mathbb{C}^{N \times N}$ and $\mathbf{F}_m \in \mathbb{C}^{M \times M}$ are Fourier matrices.
This decomposition allows us to take advantage of the \ac{FFT} for faster computation.
To this end, we define a matrix $\mathbf{R} \in \mathbb{C}^{N \times M}$, whose entries at the indices $\Omega_{\rm{s}}$ are initialized with the values of the residual vector $\mathbf{h}_r$, (i.e., $\mathbf{R}(\Omega_{\rm{s}}) = \mathbf{h}_r$).
Then, we exploit the \ac{FFT} to compute the correlation as:
\begin{equation}
    \begin{aligned}
        \mathbf{c}_{\ell} &= \left(\mathbf{I_s} (\mathbf{F}_m^\ast \otimes \mathbf{F}_n)\right)^{\mathrm{H}} \mathbf{h}_{r}\\
                     &= \text{vec}\left(
                        \mathcal{F}\left(\mathcal{F}^{-1}(\mathbf{R})\right)^{\mathrm{T}}
                    \right)
    \end{aligned}
\end{equation}
Here, $\mathcal{F}$ and $\mathcal{F}^{-1}$ represent the FFT and IFFT, respectively, and $\text{vec}(\cdot)$ vectorizes the resulting matrix.
By using (I)FFT we significantly reduce the computational complexity of the correlation step from $\mathcal{O}(|\Omega_{\rm{s}}|NM)$ (i.e., direct matrix multiplication) to $\mathcal{O}(NM(\log(N)+\log(M)))$.

Once the correlation vector $\mathbf{c}_{\ell}$ is computed, the coarse on-grid estimate $(\hat\tau_{\ell},\hat\alpha_{\ell})$ is selected as
\begin{equation}
    \begin{aligned}
        \Pi_{\ell} &=  \arg\max_j |\mathbf{c}_{\ell}(j)|^2,\\
        (\hat \tau_{\ell} ,\hat \alpha_{\ell}) &= \text{ind2sub}\left([N,M], \Pi_{\ell}\right),
\end{aligned}
\end{equation}
where the function $\text{ind2sub}$ converts linear index $\Pi_{\ell}$ to subscript indices $\hat\tau_{\ell}$ and $\hat\alpha_{\ell}$ as
\begin{equation}
    \begin{aligned} 
        \hat\tau_{\ell} &= \Pi_{\ell}- (\left \lceil{ \Pi_{\ell}/N }\right \rceil -1)N,\\ 
        \hat\alpha_{\ell} &= \left \lceil{ \Pi_{\ell}/N }\right \rceil.
\end{aligned}
\end{equation}
We obtain the corresponding attenuation coefficients $\hat\beta_{\ell}$ by invoking step 10 of \Cref{omp_algo}. 
Hereafter in all subsequent sections, we assume ${\mathbf{a}} = \mathbf{a}^{\Omega_{\rm{s}}}$ for clarity and brevity.
\subsubsection{Local Newton Refinements}\label{local_refine}

In this step, the coarse estimates $(\hat\tau_{\ell},\hat\alpha_{\ell},\hat\beta_{\ell})$ of a newly detected target are refined by the Newton method. 
The refined estimates of the coarse estimates are obtained by solving the non-linear least squares problem $\Vert\mathbf{h}_r - \mathbf{a}(\hat\tau_{\ell},\hat\alpha_{\ell} )\hat\beta_{\ell}\Vert_2^2$, which is equivalent to maximizing
\begin{equation}
    {\mathbf{S}}\left( {\hat\tau_{\ell} ,\hat\alpha_{\ell}, \hat\beta_{\ell}} \right) = 2\Re \left\{ {{\mathbf{h}}_r^{\mathrm{H}}{\mathbf{a}}(\hat\tau_{\ell},\hat\alpha_{\ell} )\hat\beta_{\ell}} \right\} - | \hat\beta_{\ell}|^2\Vert{\mathbf{a}}(\hat\tau_{\ell},\hat\alpha_{\ell} )\Vert_2^2.
\end{equation}
The Newton refinement of $(\hat\tau_{\ell},\hat\alpha_{\ell})$ is done by
\begin{equation}\label{eq:newton_iter}
    \left[ {\begin{array}{c} {{{\hat\tau_{\ell} }^\prime }} \\ {{{\hat\alpha_{\ell}}^\prime }} \end{array}} \right] = \left[ {\begin{array}{c} {\hat\tau_{\ell}} \\ {\hat\alpha_{\ell}} \end{array}} \right] - \mathbf{\ddot S}{\left( {\hat\tau_{\ell} ,\hat\alpha_{\ell}, \hat\beta_{\ell}} \right)^{ - 1}}\mathbf{\dot S}\left( {\hat\tau_{\ell} ,\hat\alpha_{\ell}, \hat\beta_{\ell}} \right),
\end{equation}
where $\mathbf{\dot S}$ is known as the Jacobian matrix and is defined as
\begin{equation}
    \mathbf{\dot S}\left( {\hat\tau_{\ell} ,\hat\alpha_{\ell}, \hat\beta_{\ell}} \right) = \left[ {\begin{array}{c} {\frac{{\partial {\mathbf{S}}}}{{\partial \hat\tau_{\ell} }}} \\ {\frac{{\partial {\mathbf{S}}}}{{\partial \hat\alpha_{\ell}}}} \end{array}} \right],\label{jacob_single}
\end{equation}
and $\mathbf{\ddot S}$ is known as the Hessian matrix and is defined as
\begin{equation}
    \mathbf{\ddot S}\left( {\hat\tau_{\ell} ,\hat\alpha_{\ell}, \hat\beta_l} \right) = \left[ {\begin{array}{cc} {\frac{{{\partial ^2}\mathbf{S}}}{{\partial {{\hat\tau_{\ell} }^2}}}}&{\frac{{{\partial ^2}\mathbf{S}}}{{\partial \hat\tau_{\ell} \partial \hat\alpha_{\ell}}}} \\ {\frac{{{\partial ^2}\mathbf{S}}}{{\partial \hat\alpha_{\ell} \partial \hat\tau_{\ell} }}}&{\frac{{{\partial ^2}\mathbf{S}}}{{\partial {{\hat\alpha_{\ell}}^2}}}} \end{array}} \right].
\end{equation}
The first order partial derivatives of $\mathbf{S}$ w.r.t $\hat\tau_{\ell}$ and $\hat\alpha_{\ell}$ are computed as
\begin{align} 
    & \frac{{\partial {\mathbf{S}}}}{{\partial \hat\tau_{\ell} }} = 2\Re \left\{ {{{\left( {{{\mathbf{h}}_r} - {\mathbf{a}}(\hat\tau_{\ell},\hat\alpha_{\ell} )\hat\beta_{\ell}} \right)}^{\mathrm{H}}}\frac{{\partial {\mathbf{a}}(\hat\tau_{\ell},\hat\alpha_{\ell} )}}{{\partial \hat\tau_{\ell} }}\hat\beta_{\ell}} \right\},\\
    & \frac{{\partial {\mathbf{S}}}}{{\partial \hat\alpha_{\ell}}} = 2\Re \left\{ {{{\left( {{{\mathbf{h}}_r} - {\mathbf{a}}(\hat\tau_{\ell},\hat\alpha_{\ell} )\hat\beta_{\ell}} \right)}^{\mathrm{H}}}\frac{{\partial {\mathbf{a}}(\hat\tau_{\ell},\hat\alpha_{\ell} )}}{{\partial \hat\alpha_{\ell} }}\hat\beta_{\ell}} \right\}.
\end{align}
Similarly, the second-order derivatives are computed as
\begin{equation}
    \begin{aligned}
        \frac{{\partial^2 \mathbf{S}}}{{\partial \hat\tau_{\ell}^2}} & = 2\Re \left\{ \left( \mathbf{h}_r - \mathbf{a}(\hat\tau_{\ell},\hat\alpha_{\ell})\hat\beta_{\ell} \right)^{\mathrm{H}} \frac{{\partial^2 \mathbf{a}(\hat\tau_{\ell},\hat\alpha_{\ell})}}{{\partial \hat\tau_{\ell}^2}} \hat\beta_{\ell}\right.\\ 
        & \quad\left. - |\hat\beta_{\ell}|^2 \frac{{\partial \mathbf{a}^{\mathrm{H}}(\hat\tau_{\ell},\hat\alpha_{\ell})}}{{\partial \hat\tau_{\ell}}} \frac{{\partial \mathbf{a}(\hat\tau_{\ell},\hat\alpha_{\ell})}}{{\partial \hat\tau_{\ell}}}\right\}, 
    \end{aligned}
\end{equation}
\begin{equation}
    \begin{aligned}
        \frac{{\partial^2 \mathbf{S}}}{{\partial \hat\alpha_{\ell}^2}} & = 2\Re \left\{ \left( \mathbf{h}_r - \mathbf{a}(\hat\tau_{\ell},\hat\alpha_{\ell})\hat\beta_{\ell} \right)^{\mathrm{H}} \frac{{\partial^2 \mathbf{a}(\hat\tau_{\ell},\hat\alpha_{\ell})}}{{\partial \hat\alpha_{\ell}^2}} \hat\beta_{\ell}\right. \\
        & \quad \left. - |\hat\beta_{\ell}|^2 \frac{{\partial \mathbf{a}^{\mathrm{H}}(\hat\tau_{\ell},\hat\alpha_{\ell})}}{{\partial \hat\alpha_{\ell}}} \frac{{\partial \mathbf{a}(\hat\tau_{\ell},\hat\alpha_{\ell})}}{{\partial \hat\alpha_{\ell}}} \right\},
    \end{aligned}
\end{equation}
\begin{equation}
    \begin{aligned}
        \frac{\partial^2 \mathbf{S}}{\partial \hat\tau_{\ell} \partial \hat\alpha_{\ell}} & = \frac{\partial^2 \mathbf{S}}{\partial \hat\alpha_{\ell} \partial \hat\tau_{\ell}} \\
        & = 2\Re \left\{ \left( \mathbf{h}_r - \mathbf{a}(\hat\tau_{\ell},\hat\alpha_{\ell})\hat\beta_{\ell} \right)^{\mathrm{H}} \frac{\partial^2 \mathbf{a}(\hat\tau_{\ell},\hat\alpha_{\ell})}{\partial \hat\alpha_{\ell}\partial \hat\tau_{\ell}} \hat\beta_{\ell} \right. \\
        & \quad \left. -|\hat\beta_{\ell}|^2 \frac{\partial \mathbf{a}^{\mathrm{H}}(\hat\tau_{\ell},\hat\alpha_{\ell})}{\partial \hat\alpha_{\ell}} \frac{\partial \mathbf{a}(\hat\tau_{\ell},\hat\alpha_{\ell})}{\partial \hat\tau_{\ell}} \right\}.
    \end{aligned}
\end{equation}
When plugging these into \eqref{eq:newton_iter} and iterating it, we obtain the locally refined delay ${\hat\tau_{\ell} }^\prime$ and Doppler shift ${\hat\alpha_{\ell}}^\prime$ estimates.
Then we can update the target's attenuation coefficient $\hat\beta_{\ell}^\prime$ at the end of each Newton refinement step via
\begin{equation}
    \hat\beta_{\ell}^{\prime} = {{\mathbf{a}}^{\mathrm{H}}}\left( {{{\hat\tau_{\ell}}^{\prime}},{{\hat\alpha_{\ell} }^{\prime}}} \right){{\mathbf{h}}_r}/\Vert {{\mathbf{a}}\left( {{{\hat\tau_{\ell}}^{\prime}},{{\hat\alpha_{\ell}}^{\prime}}} \right)} \Vert_2^2.
\end{equation}
Finally, the residual vector $\mathbf{h}_r$ will be updated according to \eqref{residual}.
This step is repeated $R_s$ times, where $R_s$ is the number of Newton refinements.
\subsubsection{Global Newton Refinements}\label{global_refine}

After the Newton refinements of individual targets' parameters, the estimated refined parameters set is $\mathcal{C}^\prime = \{\left(\hat\tau_{\ell}^\prime,\hat\alpha_{\ell}^\prime,\hat\beta_{\ell}^\prime\right)\mid {\ell}=1,\cdots,K\}$. 
In this step, all the refined estimates $\left(\tau_{\ell}^\prime,\alpha_{\ell}^\prime,\beta_{\ell}^\prime\right)$ in $\mathcal{C}^\prime$ are \emph{jointly} fine-tuned via the joint Newton refinements.
Consider $K^\text{th}$ round, where $K$ number of targets have been discovered.
At this point, we jointly fine-update the targets' estimates using the following optimization problem.
\begin{equation}
    \begin{aligned}
        \mathbf{S}(\mathcal{C}^\prime) &= \Vert\mathbf{h}_r - \sum\limits_{{\ell} = 1}^{K}{\hat\beta_{\ell}^\prime}\mathbf{a}\left( {{\hat\tau_{\ell}^\prime},{\hat\alpha_{\ell}^\prime}} \right)\Vert_2^2\\
        &= \Vert\mathbf{h}_r - \mathbf{A}_{\mathcal{C}^\prime}\mathbf{b}_{\mathcal{C}^\prime}\Vert_2^2.
    \end{aligned}
\end{equation}
This problem can be solved by the joint Newton refinement method as given below.
\begin{equation}
    {\begin{bmatrix}
        {{{\hat\tau_1 }^{\prime\prime} }},\cdots,{{{\hat\tau_K }^{\prime\prime} }} \\ {{{\hat\alpha_1}^{\prime\prime} }},\cdots,{{{\hat\alpha_K}^{\prime\prime} }} \end{bmatrix}} =  {\begin{bmatrix} {\hat\tau_1^\prime },\cdots,{\hat\tau_K^\prime } \\ {\hat\alpha_1^\prime},\cdots,{\hat\alpha_K^\prime}
    \end{bmatrix}} - \mathbf{\ddot S}(\mathcal{C}^\prime)^{-1}\mathbf{\dot S}(\mathcal{C}^\prime),\label{joint_newton}
\end{equation}
where, $\mathbf{\ddot S}(\mathcal{C}^\prime)\in \mathbb{R}^{2K \times 2K}$ and $\mathbf{\dot S}(\mathcal{C}^\prime) \in \mathbb{R}^{2K \times 1}$ are block Hessian and block Jacobian matrices, which are given below.
\begin{equation}
   \mathbf{\ddot S}(\mathcal{C}^\prime)=\left[ {\begin{array}{cccc} {{{\mathbf{\ddot S}}_{11}}}&{{{\mathbf{\ddot S}}_{12}}}& \cdots &{{{\mathbf{\ddot S}}_{1K}}} \\ {{{\mathbf{\ddot S}}_{21}}}&{{{\mathbf{\ddot S}}_{22}}}& \cdots &{{{\mathbf{\ddot S}}_{2K}}} \\ \vdots & \vdots & \ddots & \vdots \\ {{{\mathbf{\ddot S}}_{K1}}}&{{{\mathbf{\ddot S}}_{K2}}}& \cdots &{{{\mathbf{\ddot S}}_{KK}}} \end{array}} \right], \mathbf{\dot S}(\mathcal{C}^\prime)=\begin{bmatrix}
        \mathbf{\dot S}_1\\
        \vdots\\
        \mathbf{\dot S}_K
    \end{bmatrix}.\label{blk_Hess_Jacob}
\end{equation}
The expressions of the entries of the matrix $\mathbf{\ddot S}(\mathcal{C}^\prime)$, and the matrix $\mathbf{\dot S}(\mathcal{C}^\prime)$ are rather lengthy and can be found in the Appendix.

Notably, due to the block hessian matrix structure, the computational complexity of the joint Newton refinement step in \eqref{joint_newton} is $\mathcal{O}\left((3K)^3\right)$ which increases non-linearly with the number of targets. 
Thus, we propose low-complexity update by relaxing the block Hessian matrix to a block diagonal matrix as
\begin{equation}
   \mathbf{\ddot S}(\mathcal{C}^\prime)=\left[ {\begin{array}{cccc} {{{\mathbf{\ddot S}}_{11}}}&{\mathbf{0}}& \cdots &{\mathbf{0}} \\ {\mathbf{0}}&{{{\mathbf{\ddot S}}_{22}}}& \cdots &{\mathbf{0}} \\ \vdots & \vdots & \ddots & \vdots \\ {\mathbf{0}}&{\mathbf{0}}& \cdots &{{{\mathbf{\ddot S}}_{KK}}} \end{array}} \right],
\end{equation}
This reduces the complexity to $\mathcal{O}\left(K\right)$ per step.
Finally the attenuation coefficients vector $\mathbf{b}_{\mathcal{C}^{\prime\prime}}$ corresponding to the updated parameters set $\mathcal{C}^{\prime\prime} = \{\left(\hat\tau_{\ell}^{\prime\prime},\hat\alpha_{\ell}^{\prime\prime},\hat\beta_{\ell}^{\prime\prime}\right)\mid {\ell}=1,\cdots,K\}$ is obtained by least squares method as
\begin{equation}
    \mathbf{b}_{\mathcal{C}^{\prime\prime}} = \mathbf{A}_{\mathcal{C}^{\prime\prime}}^\dagger\mathbf{h}_s,
\end{equation}
where, $\mathbf{A}_{\mathcal{C}^{\prime\prime}}\triangleq\left[\mathbf{a}({{{\hat \tau_1 }^{\prime\prime} }},{{{\hat \alpha_1 }^{\prime\prime} }}),\cdots,\mathbf{a}({{{\hat \tau_K }^{\prime\prime} }},{{{\hat \alpha_K }^{\prime\prime} }})\right]$.

The algorithm terminates when $\max_j |\mathbf{c}_{\ell}(j)|^2\leq\delta$, where $\delta\triangleq\ln{|\Omega_{\rm{s}}|}-\ln{(-\ln{(1-p_{fa})})}$ is a CFAR-based threshold, which is selected based on the required probability of false alarm $p_{fa}$~\cite{Mamandipoor2016NewtonizedContinuum}.

The proposed \ac{NOMP}-based algorithm is stated in \cref{NOMP_algo}.
\begin{algorithm}
\caption{Proposed \ac{NOMP}-based Algorithm}\label{NOMP_algo}
\begin{algorithmic}[1]
\State \textbf{Input:} $\mathbf{h}_s$, $\mathbf{\bar{A}}$, $\delta$
\State \textbf{Initial state:} $\mathbf{h}_r = \mathbf{h}_s$, $\Pi_0 = \emptyset$, ${\ell} = 0$, $\mathbf{R}^0(\Omega_{\rm{s}}) = \mathbf{h}_r$
\Repeat
    \State ${\ell} = {\ell} + 1$
    \State \textbf{Initial Coarse Estimation:}
    \State \quad $\mathbf{c}_{\ell} =  \text{vec}\left(\mathcal{F}\left(\mathcal{F}^{-1}(\mathbf{R}^{{\ell}-1})\right)^{\mathrm{T}}\right)$\Comment{Faster computation}
    \State \quad $\Pi_{\ell} = \arg\max_j |\mathbf{c}_{\ell}(j)|^2$
    \State \quad $(\hat\tau_{\ell},\hat\alpha_{\ell})=\text{ind2sub}(\text{size}(\mathbf{R}), \Pi_{\ell})$
    \State \quad $\hat\beta_{\ell} = {{\mathbf{a}}^{\mathrm{H}}}\left( {{{\hat\tau_{\ell}}},{{\hat\alpha_{\ell} }}} \right){{\mathbf{h}}_r}/\Vert {{\mathbf{a}}\left( {{{\hat\tau_{\ell}}},{{\hat\alpha_{\ell}}}} \right)} \Vert_2^2$
    \State \textbf{Local Newton Refinement:}\Comment{Iterate $R_s$ times}
    \State \quad  $(\hat\tau_{\ell}^\prime,\hat\alpha_{\ell}^\prime) = (\hat\tau_{\ell},\hat\alpha_{\ell}) - \mathbf{\ddot S}^{ - 1}\mathbf{\dot S}$
    \State \quad $ \hat\beta_{\ell}^{\prime} = {{\mathbf{a}}^H}\left( {{{\hat\tau_{\ell}}^{\prime}},{{\hat\alpha_{\ell} }^{\prime}}} \right){{\mathbf{h}}_r}/\Vert {{\mathbf{a}}\left( {{{\hat\tau_{\ell}}^{\prime}},{{\hat\alpha_{\ell}}^{\prime}}} \right)} \Vert_2^2$
    \State \quad ${{\mathbf{h}}_r} = {\mathbf{h}_s} -{\hat\beta_{\ell}^\prime}\mathbf{a}\left( {{\hat\tau_{\ell}^\prime},{\hat\alpha_{\ell}^\prime}} \right)$
    \State \quad $\mathcal{C}^\prime = \mathcal{C}^\prime\cup({{\hat\tau_{\ell}^\prime},{\hat\alpha_{\ell}^\prime}, \hat\beta_{\ell}^\prime})$\Comment{Local refined estimates}
    \State \textbf{Global Newton Refinement:}
    \State \quad  $(\hat\tau_{\ell}^{\prime\prime},\hat\alpha_{\ell}^{\prime\prime})=(\hat\tau_{\ell}^\prime,\hat\alpha_{\ell}^\prime)-\mathbf{\ddot S}(\mathcal{C}^\prime)^{-1}\mathbf{\dot S}(\mathcal{C}^\prime)$\Comment{$ 1\leq {\ell}\leq K$}
    \State \quad $\mathbf{b}_{\mathcal{C}^{\prime\prime}} = \mathbf{A}_{\mathcal{C}^{\prime\prime}}^\dagger\mathbf{h}_s$\Comment{Least squares gain update}
    \State \quad $\mathbf{h}_r = \mathbf{h}_s - \mathbf{A}_{\mathcal{C}^{\prime\prime}}\mathbf{b}_{\mathcal{C}^{\prime\prime}}$\Comment{Update the residue}
    \State \quad $\mathbf{R}^{\ell}(\Omega_{\rm{s}}) = \mathbf{h}_r$\Comment{Update the $\mathbf{R}$ matrix}
    \State \quad $\Pi_{\ell} = \Pi_{{\ell}-1}\cup \text{sub2ind}(\text{size}(\mathbf{R}), {{\hat\tau_{\ell}^{\prime\prime}},{\hat\alpha_{\ell}^{\prime\prime}}})$
\Until{$\max |\mathbf{c}_{\ell}|^2\leq\delta$}
\State \textbf{Output:} $\mathbf{b}_{\mathcal{C}^{\prime\prime}}$, $\Pi$
\end{algorithmic}
\end{algorithm}
\subsection{Computational Complexity Analysis}
In this subsection, we analyze the computational complexity of the proposed \ac{NOMP}-based algorithm.
The \cref{NOMP_algo} has three major parts: Initial coarse estimation (steps 6-9), Local Newton refinement (steps 11–14), and Global Newton refinement (steps 16-20), and the computational complexity of each part is $\mathcal{O}(NM(\log(N)+\log(M)))$, $\mathcal{O}(kR_s|\Omega_{\rm{s}}|)$, and $\mathcal{O}(|\Omega_{\rm{s}}|k^2+k^3)$, respectively, where $k$ represents the current number of iterations.
\section{Simulation Results}\label{simulation}
In this section, we evaluate the performance of the proposed algorithm and compare it with conventional \ac{2D-FFT}, \ac{SALSA}~\cite{Knill2018HighRadar}, and \ac{OMP} algorithm.
We simulate an \ac{ISAC} scenario consisting of a TX vehicle, RX vehicle, and a variable number of targets (i.e., vehicles and pedestrians) moving on a road.
The TX and RX vehicles are equipped with single TX and single RX omnidirectional antennas respectively, for half-duplex communication and bi-static radar sensing.
The sparse \ac{OFDM} grid parameters used by the TX and RX vehicles, and other simulation parameters are shown in \cref{tab:settings}.

\begin{table}[htbp]
\caption{Simulation Parameters}
\label{tab:settings}
\centering
\begin{tabular}{|l|l|l|}
\hline
\multicolumn{1}{|c|}{\textbf{Symbol}} & \multicolumn{1}{c|}{\textbf{Parameter}} & \multicolumn{1}{c|}{\textbf{Value}} \\ \hline
$f_c$                                    & Carrier frequency                       & \SI{5.9}{\giga\hertz}                             \\ \hline
$N$                                     & Total subcarriers                   & $1560$                                \\ \hline
$n\in\Omega_{\rm{s}}$                                    & Subcarriers used per OFDM symbol              & $78$                                  \\ \hline
$M$                                     & Total OFDM symbols                  & $280$                                 \\ \hline
$m\in\Omega_{\rm{s}}$                                    & OFDM symbols used in OFDM grid             & $56$                                  \\ \hline
$\eta$                   & Resource occupancy                      & \SI[round-precision=2]{1}{\percent}                                \\ \hline
$B$                                     & Bandwidth                              & \SI{46.8}{\mega\hertz}                            \\ \hline
$\Delta f$               & Subcarrier spacing                      & \SI{30}{\kilo\hertz}                              \\ \hline
$K$                                     & Number of targets               & variable                            \\ \hline
$R_s$                                    & Newton refinement steps                 & $5$                                   \\ \hline
\end{tabular}
\end{table}

\begin{figure}[htbp]
    \centering
    \includegraphics[width=3in,height=\textwidth,keepaspectratio]{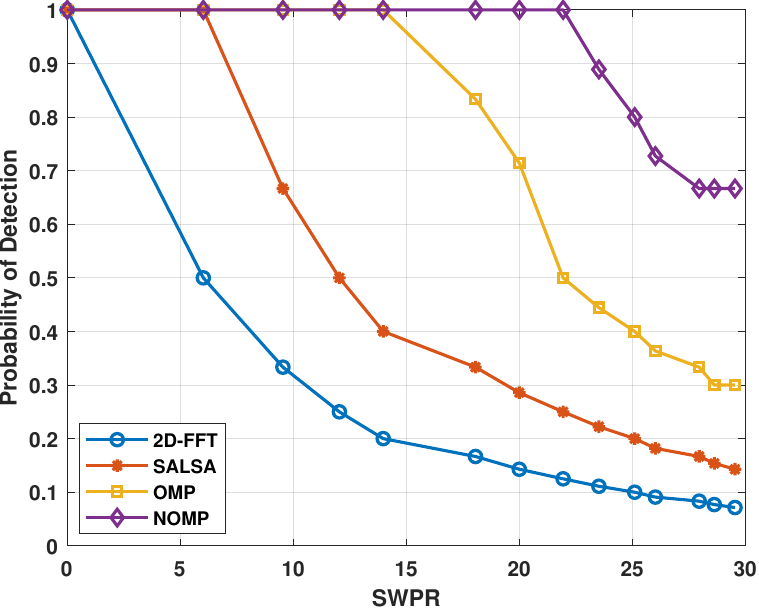}
    \caption{Probability of weak target detection in the presence of strong target}
    \label{fig:PoD}
\end{figure}

To show the ability of the proposed method to detect weak targets (e.g., pedestrians) in the presence of strong targets (e.g., vehicles) and \ac{AWGN}, we evaluate the probability of detection as a function of \ac{SWPR} as shown in \cref{fig:PoD}.
The larger values of \ac{SWPR} implies lower peak power of weak targets.
The \ac{2D-FFT} shows worst performance followed by \ac{SALSA} and \ac{OMP}.
The \ac{SALSA} and \ac{OMP} fully recovers weak targets up to \ac{SWPR} of about \SI[round-precision=2]{6}{\decibel} and \SI{14}{\decibel} respectively.
The \ac{NOMP}-based algorithm outperforms all approaches with perfect recovery of weak targets upto \ac{SWPR} of about \SI{23}{\decibel}.

\begin{figure}[htbp]
    \centering
    \includegraphics[width=3in,height=\textwidth,keepaspectratio]{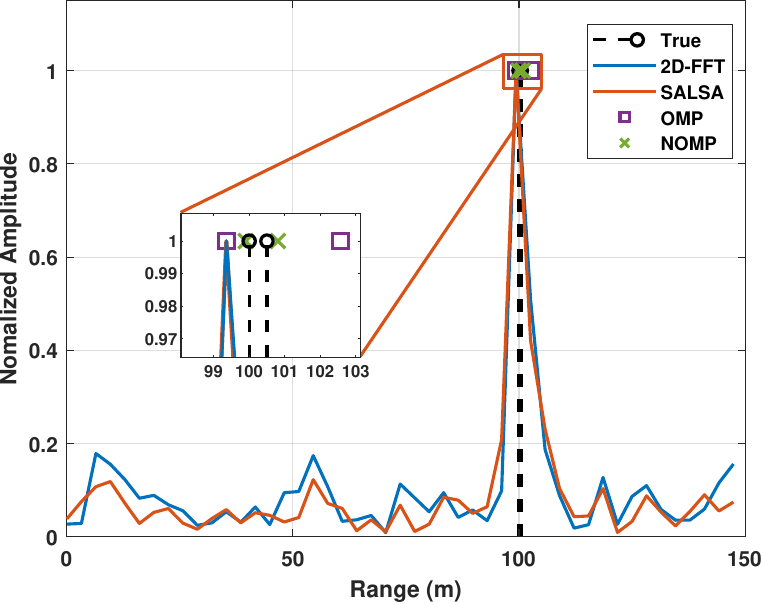}
    \caption{Recovery of two closely spaced off-grid targets in range domain}
    \label{fig:close_range}
\end{figure}

\begin{figure}[htbp]
    \centering
    \includegraphics[width=3in,height=\textwidth,keepaspectratio]{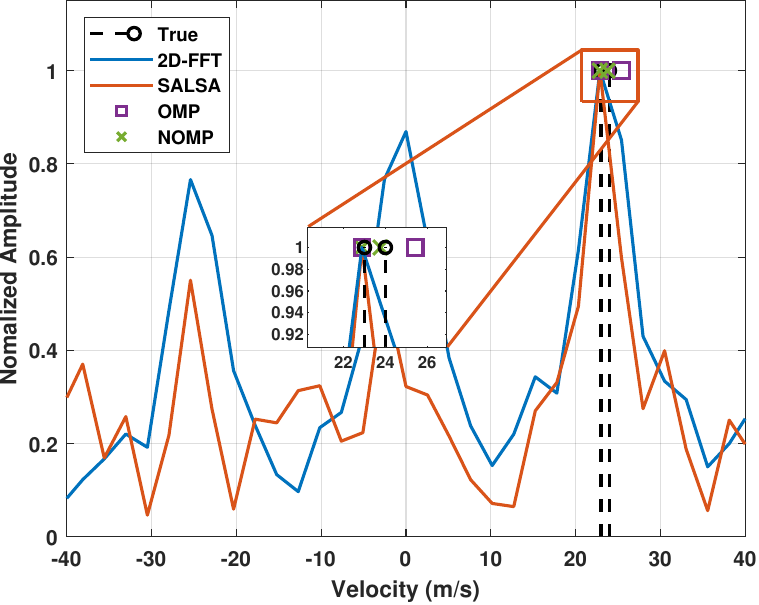}
    \caption{Recovery of two closely spaced off-grid targets in velocity domain}
    \label{fig:close_vel}
\end{figure}

Next, we evaluate the ability to distinguish two closely spaced off-grid targets in both the range and velocity domain as shown in \cref{fig:close_range} and \cref{fig:close_vel} respectively.
In the range domain, the two targets are separated by \SI[round-precision=2]{0.5}{\meter} (i.e., one is kept at \SI{100}{\meter} while the other is placed at \SI[round-precision=4]{100.5}{\meter}) as shown in the maximized region of \cref{fig:close_range}.
The \ac{2D-FFT} and \ac{SALSA} show similar performance and can recover only one target with a peak at the grid point close to the true range.
The \ac{OMP} detects both targets but with inadequate accuracy (i.e., the peak of the second target is more than \SI[round-precision=2]{2}{\meter} away from the true range).
The \ac{NOMP} not only successfully distinguishes two close targets but also accurately estimates the off-grid ranges.
On the other hand, in the velocity domain, the two targets are separated by \SI[round-precision=2]{1}{\meter\per\second} (i.e., one is moving at \SI{23}{\meter\per\second} while the other is at \SI{24}{\meter\per\second}) as shown in \cref{fig:close_vel}.
Here, the \ac{2D-FFT} and \ac{SALSA} show the worst performance with sidelobes comparable in magnitude to the main lobe (e.g, \ac{PSLR} of around \SI{1.5}{\decibel} is observed).
This behavior can lead to false target detections and reduced overall detection reliability and accuracy.
Moreover, both algorithms fail to resolve the targets in the velocity domain and produce a single merged main lobe appearing at the nearest grid point to one of the true target velocities.
The \ac{OMP} again shows two peaks, one is quite accurate while the other has a deviation of around \SI[round-precision=2]{2}{\meter\per\second}.
The \ac{NOMP} outperforms the \ac{OMP} by estimating the off-grid velocities with minimum deviations.

Then we investigate the estimation accuracies of the aforementioned algorithms as a function of SNR.
We evaluate the \ac{RMSE} in the range and velocity estimation for a single target case ($K\,=1$) by running $500$ independent \ac{MC} simulations for each SNR.
In each \ac{MC} iteration, the subset $\Omega_{\rm{s}}$ of resources is randomly selected to form sparse TX matrix $\mathbf{X}$, and the noise matrix $\mathbf{Z}$ is also randomly generated, following Gaussian distribution.
The target's range and velocity are randomly chosen from $\mathcal{U}(\SI{50}{\meter}, \SI{1500}{\meter})$ and $\mathcal{U}(\SI{-30}{\meter\per\second}, \SI{30}{\meter\per\second})$ respectively.
The \ac{CRB} given in~~\cite{Keskin2021LimitedRadar-Communications} is taken as a benchmark.
The \acp{CRB} for range and velocity estimation are expressed as follows:
\begin{equation}
    \begin{aligned}
        \sigma_{range}^2 \geq& \frac{3c^{2}}{\mathsf{SNR}8\pi^{2}MN(n^{2}-1)\Delta f^{2}},\\
        \sigma_{velocity}^2 \geq& \frac{3c^{2}}{\mathsf{SNR}8\pi^{2}f_{\mathrm{c}}^{2}MN(m^{2}-1)T_{o}^{2}}.
    \end{aligned}
\end{equation}

\begin{figure}[htbp]
    \centering
    \includegraphics[width=3in,height=\textwidth,keepaspectratio]{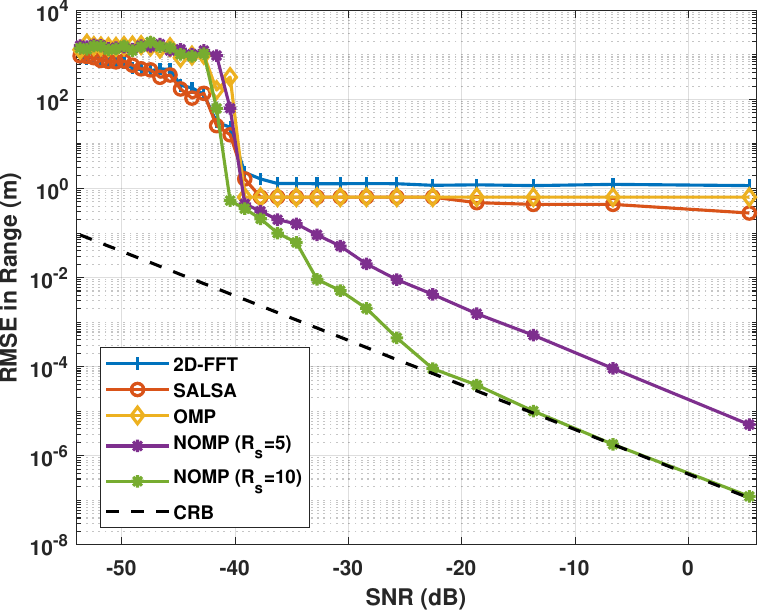}
    \caption{RMSE in range estimation}
    \label{fig:rmse_range}
\end{figure}

\begin{figure}[htbp]
    \centering
    \includegraphics[width=3in,height=\textwidth,keepaspectratio]{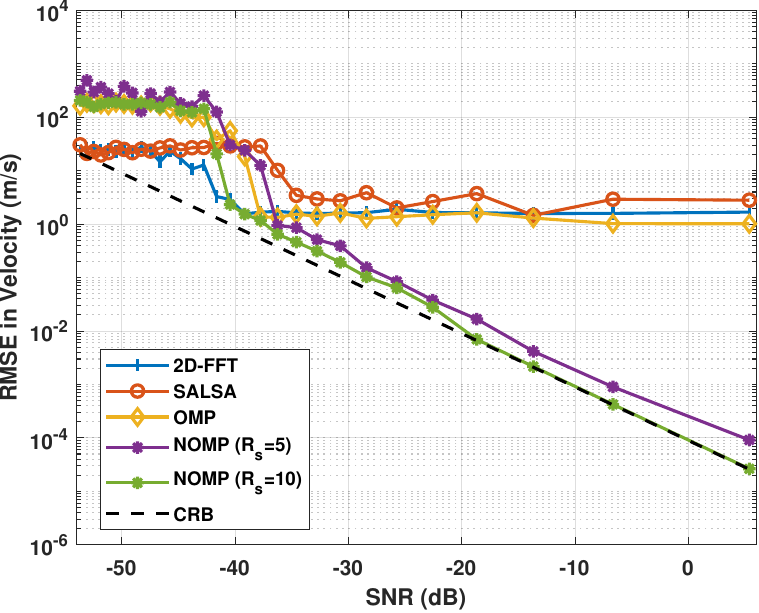 }
    \caption{RMSE in velocity estimation}
    \label{fig:rmse_vel}
\end{figure}

The \acp{RMSE} in range and velocity estimation are shown in \cref{fig:rmse_range} and \cref{fig:rmse_vel} respectively, which show that the proposed \ac{NOMP}-based algorithm not only outperforms the other three approaches but also gradually approaches the \ac{CRB}.
When increasing the Newton refinements from $R_s\,=5$ to $R_s\,=10$, the \ac{RMSE} range and velocity curves converge faster and meet the \ac{CRB} curve.
Thus a trade-off exists between accuracy requirements and complexity.
On the other hand, since the \ac{2D-FFT}, \ac{SALSA}, and \ac{OMP} support on-grid target search, their \acp{RMSE} (i.e., in both range and velocity estimation) no longer improve with the increase in SNR.

\begin{figure}[htbp]
    \centering
    \includegraphics[width=3in,height=\textwidth,keepaspectratio]{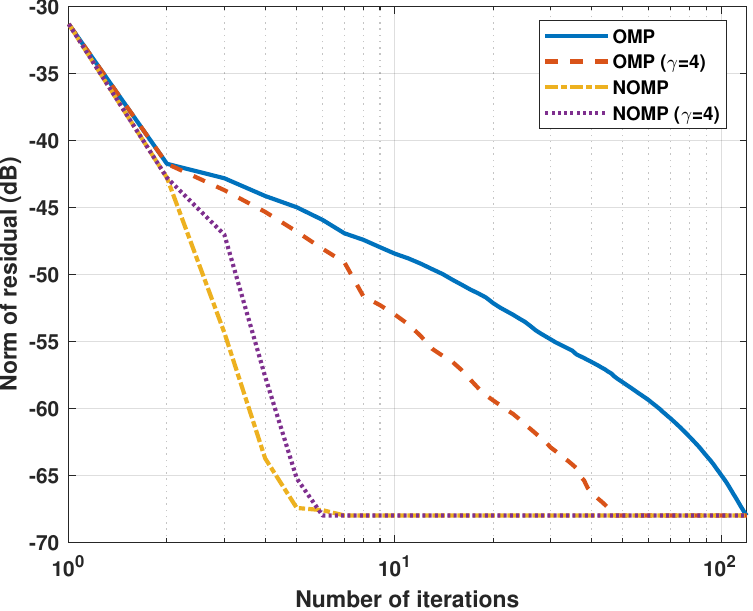}
    \caption{Convergence rates of OMP and NOMP with and without over-sampled grid at $K\,=6$}
    \label{fig:converge}
\end{figure}

In \cref{fig:converge}, we illustrate the convergence rate of the \ac{NOMP} and compare it with that of the \ac{OMP}.
We analyze the effect of the oversampling factor ($\gamma$) on the convergence rate.
By setting $\gamma = 4$. and $K\,=6$, we simulate the norm of the residual as a function of the number of iterations.
The results show that the \ac{NOMP}, both with and without oversampling, converges much faster than the \ac{OMP} in both cases.
In fact, the \ac{NOMP} reduces the residual energy more quickly by utilizing the local and joint Newton refinements steps.
The \ac{NOMP} with oversampling converges after $6$ iterations (equal to $K$) while \ac{NOMP} without oversampling achieves convergence after $7$ iterations estimating one spurious ghost target.
The convergence times for the \ac{NOMP} with and without oversampling are \SI{3.95}{\second}and \SI{1.03}{\second}, respectively.
In contrast, the \ac{OMP} takes \SI{195.36}{\second} with the oversampling factor and \SI{56.15}{\second} without it.
These simulations were conducted on a \SI{3}{\giga\hertz} Intel(R) Core(TM) i5-7400 CPU.

\begin{figure}[htbp]
    \centering
    \includegraphics[width=3in,height=\textwidth,keepaspectratio]{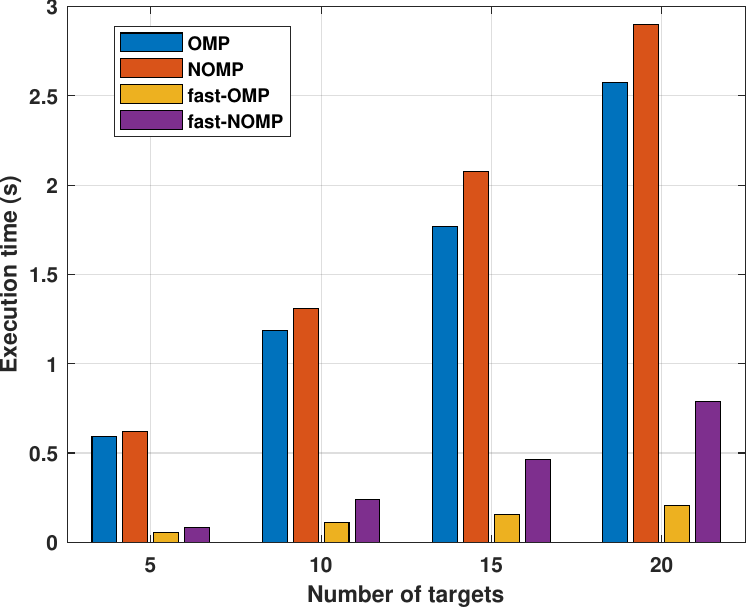}
    \caption{Execution time per iteration comparison of \ac{OMP}, \ac{NOMP}, and their faster implementations as a function of number of targets}
    \label{fig:exc_time}
\end{figure}

In \cref{fig:exc_time}, we show the execution time per iteration of the \ac{OMP}, \ac{NOMP}, and their faster implementations.
Both \ac{OMP} and \ac{NOMP} compute the correlations between the residual and the columns of the sensing matrix in each iteration, which has a computational complexity of $\mathcal{O}\left(\gamma |\Omega_{\rm{s}}|(MN)\right)$.
By replacing this correlation step with a more efficient \ac{2D-FFT} operation, the complexity of the algorithms can be reduced to $\mathcal{O}\left(\gamma |\Omega_{\rm{s}}|\log(MN)\right)$, leading to faster implementations.
\section{Measurement Setup}\label{measurement}
To validate the proposed algorithm, we employ a bi-static radar setup in a semi-anechoic chamber, namely the Virtual Road Simulation and Test Area (VISTA), at the Thuringian Center of Innovation in Mobility.
VISTA features a metal-shielded semi-anechoic chamber with inner dimensions of $\SI{13}{\meter} \times \SI[round-precision=2]{9}{\meter} \times \SI[round-precision=2]{7.5}{\meter}$.
This chamber is equipped with a \SI{6.5}{\meter} diameter turntable and an \SI[round-precision=2]{8}{\meter} diameter spherical near-field antenna measurement system.
This state-of-the-art test facility supports various RF measurement setups, including automotive antenna measurements, over-the-air channel emulation, and automotive radar testing~\cite{Hein2015EmulationSystems}.
To conduct bi-static dynamic scattering measurements for predefined traffic reference scenarios under reproducible laboratory conditions, it is essential to cover the full range of bi-static angles from \SI{0}{\degree} to \SI{360}{\degree}, as well as the complete spectrum of bi-static Doppler frequencies, which vary based on the target velocity~\cite{Schwind2020BistaticApplications}.

\subsection{Bi-static Radar Configuration}
\begin{figure}[htbp]
    \centering
    \includegraphics[width=3in,height=\textwidth,keepaspectratio]{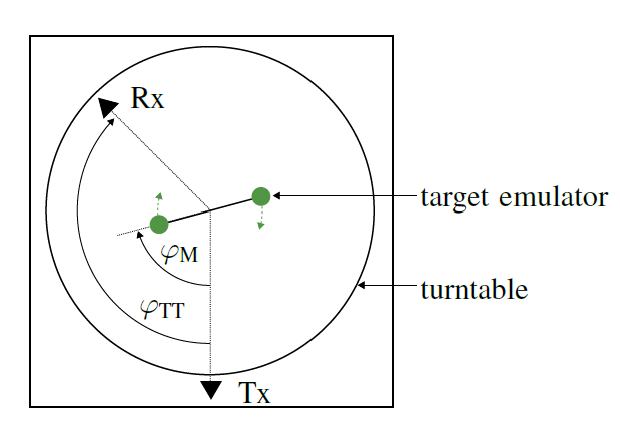}
    \caption{Bi-static radar configuration with metallic spheres emulating rotating targets}
    \label{fig:bi-static-config}
\end{figure}

\cref{fig:bi-static-config} depicts the bi-static setup, with the transmitter placed outside the turntable and the receiver mounted on it.
By positioning the radar target at the rotational center of the turntable, this configuration allows the bi-static angle to be varied across the full \SI{360}{\degree} azimuth range.
In the context of stationary radar networks, the Doppler effect observed during measurements is a result of the motion of radar targets.
To effectively validate the estimation outcomes, it is essential that this movement can be described analytically, thereby providing the necessary ground truth data.
This approach ensures that the measurement setup adequately simulates real-world conditions and allows for precise evaluation of the radar system's performance.
\subsection{Target Emulator and Measurement Hardware}\label{target_emulator}
\begin{figure}[htbp]
    \centering
    \includegraphics[width=3in,height=\textwidth,keepaspectratio]{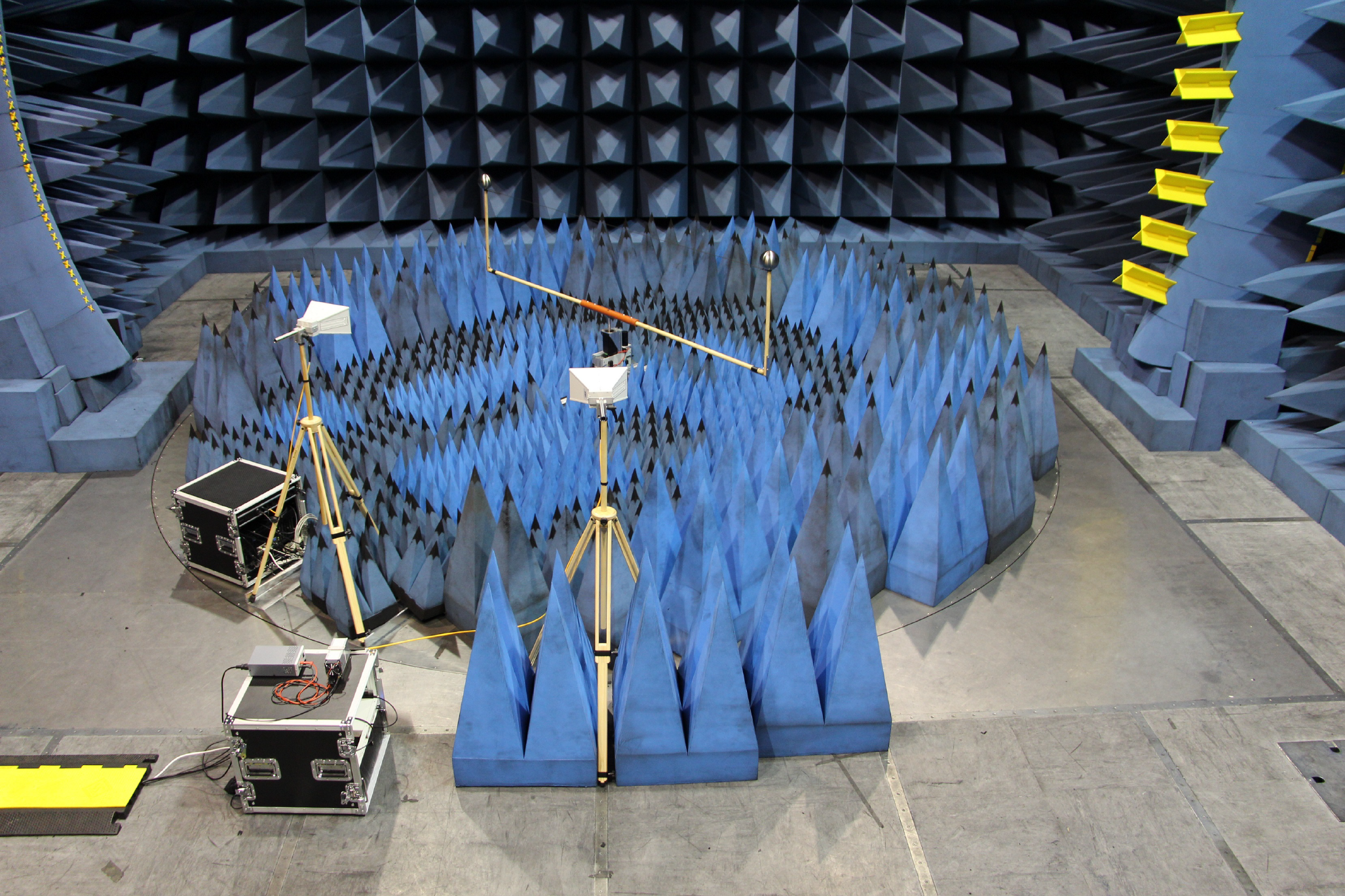}
    \caption{Photograph of the bi-static test setup at VISTA~\cite{Dobereiner2019JointMeasurements}. The metallic spheres on the target emulators, and the receiver are mounted on the turntable, while the transmitter is placed outside the turntable. Pyramidal-shaped cones are microwave absorbers that are used to minimize ground reflections.}
    \label{fig:vista_measurement_setup}
\end{figure}

The objective of the target emulator is to replicate Doppler frequencies typical of urban traffic scenarios within a straightforward and reproducible test setup.
This approach allows the signal processing software and its algorithms to be tested and evaluated under well-controlled laboratory conditions before moving to more resource-intensive field tests.
\cref{fig:vista_measurement_setup} illustrates the measurement setup, which includes the carousel-type bi-static Doppler target emulator.
The VISTA facility features a turntable used to vary the angle between the transmitter and receiver. The transmit antenna is placed outside the turntable while the receive antenna is on the turntable.
A motor equipped with an absolute angle value encoder ensures constant target velocities and provides continuous knowledge of the target positions.
Two wooden cantilever beams with a span of \SI[round-precision=2]{3}{\meter} are attached to the motor's rotational center, with metallic spheres mounted at the ends to serve as radar targets.
The Doppler target emulator generates reproducible experimental data by ensuring strict synchronization between the RF measurements and the time-varying speeds and angles of the rotating cantilever beams.
Two software-defined radio devices from the National Instruments, namely Universal Software Radio Peripheral (USRP) 2954R are utilized to feed the spatially separated transmitter and receiver.
These devices facilitate precise time synchronization with \SI{10}{\mega\hertz} and $1$ pulse per second signals shared between the transmitter and receiver.
USRPs lack internal data storage capabilities and instead stream the measured data samples to and from a server connected via Ethernet.
Since the sampling clocks are derived from the \SI{10}{\mega\hertz} signal, all samples on the Rx and Tx are inherently time-stamped.
In addition to data streaming, the Rx server controls the VISTA turntable and the motor of the radar target emulator.
\subsection{Measurement Parameters}
To maximize the spatial resolution of radar-based range measurements, the USRPs are configured to operate at their maximum analog frequency bandwidth of \SI{160}{\mega\hertz}, centered at \SI{5.9}{\giga\hertz}.
The radar signals used in the measurements are typical \ac{OFDM} symbols of LTE.
The wideband transmit signal, a Newman sequence, is continuously transmitted with a period of \SI{64}{\micro\second} similar to LTE symbol duration (i.e. \SI{66.7}{\micro\second}).

We conduct measurements of $4$ different setups. Each setup has a specific combination of sphere diameter and angular speed as shown in \cref{tab:measure-setups}.
\begin{table}[htbp]
\caption{Radar Target Parameters}
\label{tab:measure-setups}
\centering
\begin{tabular}{|l|l|l|}
\hline
\multicolumn{1}{|c|}{\textbf{Setup}} & \multicolumn{1}{c|}{\textbf{Sphere diameter}} & \multicolumn{1}{c|}{\textbf{Angular speed}} \\ \hline
Setup 1                                    & \SI[round-precision=2]{6}{\centi\meter}                       & \SI{60}{\mathrm{rpm}}                             \\ \hline
Setup 2                                     & \SI[round-precision=2]{6}{\centi\meter}                   & \SI{80}{\mathrm{rpm}}                                 \\ \hline
Setup 3                                   & \SI{12}{\centi\meter}              & \SI{30}{\mathrm{rpm}}                                  \\ \hline
Setup 4                                     & \SI{12}{\centi\meter}                  & \SI{60}{\mathrm{rpm}}                                 \\ \hline
\end{tabular}
\end{table}
Setups 1 and 2 involve small-sized spheres (\SI[round-precision=2]{6}{\centi\meter} diameter) rotating at two different speeds: \SI{60}{\mathrm{rpm}} (slow) and \SI{80}{\mathrm{rpm}} (fast) respectively.
On the other hand, setups 3 and 4 feature large-sized spheres (\SI{12}{\centi\meter} diameter) rotating at speeds of \SI{30}{\mathrm{rpm}} (slow) and \SI{60}{\mathrm{rpm}} (fast) respectively.
The maximum rotation speed of \SI{80}{\mathrm{rpm}} ensures tangential velocities of the spheres up to \SI{45}{\kilo\meter\per\hour}, which is comparable to realistic vehicle speed in urban traffic.
To emulate various combinations of bi-static delay and Doppler shift, we measure a full rotation of the turntable in \SI[round-precision=2]{5}{\degree} increments for each setup.
During the data recording, the turntable remains stationary while the motor operates at a constant speed.
This allows comprehensive evaluations of the \ac{NOMP} algorithm under controlled and reproducible conditions.
\subsection{Data Processing and Parameter Estimation}
The data processing and parameter estimation start with the time-variant frequency response of the channel matrix $\mathbf{H}$, obtained through the element-wise division between the measured received frequency spectrum $\mathbf{Y}$ and the predefined transmit frequency spectrum $\mathbf{X}$.
We select a coherent block of the first $200$ symbols from $\mathbf{H}$, which we refer to as the $\mathbf{H}_c$ matrix.
This block of $200$ symbols results in an observation time of \SI{12.8}{\milli\second}, providing a Doppler resolution of \SI[round-precision=5]{78.125}{\hertz}.
Given the frequency bandwidth of \SI{160}{\mega\hertz}, a delay resolution of \SI{6.25}{\nano\second} is achieved.
Next, exponential averaging-based background subtraction~\cite{PiccardiBackgroundReview} is applied to $\mathbf{H}_c$ to eliminate contributions from static objects, which exhibit zero Doppler shifts.
This step ensures that only the dynamic components, indicative of moving spheres, remain in the processed data.
Subsequently, we construct an unstructured sparse matrix by randomly selecting a subset $\Omega_{\rm{s}}$ of time and frequency samples from the $\mathbf{H}_c$ matrix.
By excluding unused samples from the $\mathbf{H}_c$ matrix, we compress it to a sparse matrix $\mathbf{H}_s$.
For the parameter estimation, $\mathbf{H}_s$ and its corresponding sensing matrix are input to the \ac{NOMP} algorithm.
We repeat this process for each consecutive block of 200 coherent symbols using the same set of resources $\Omega_{\rm{s}}$ until the last block to assess the performance of the \ac{NOMP} algorithm for every degree of motor rotation over a complete rotation.
These processing steps ensure a detailed evaluation of the \ac{NOMP}'s efficacy in estimating delays and Doppler shifts in sparse \ac{OFDM} \ac{ISAC} systems, providing validation of the proposed \ac{NOMP}-based target detection method through experimental results.
\subsection{Measurement Results}
\begin{figure*}
\centering
  \subcaptionbox{Setup 1}{\includegraphics[width=3in,height=\textwidth,keepaspectratio]{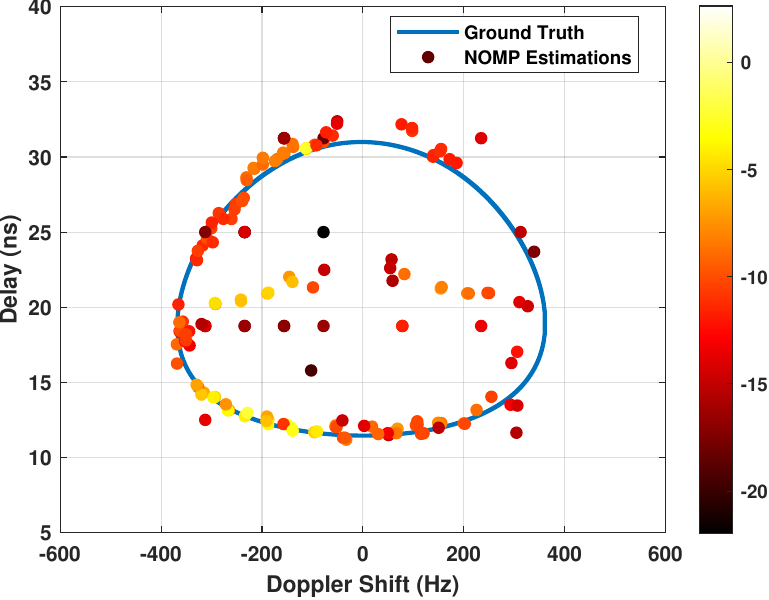}}\quad
  \subcaptionbox{Setup 2}{\includegraphics[width=3in,height=\textwidth,keepaspectratio]{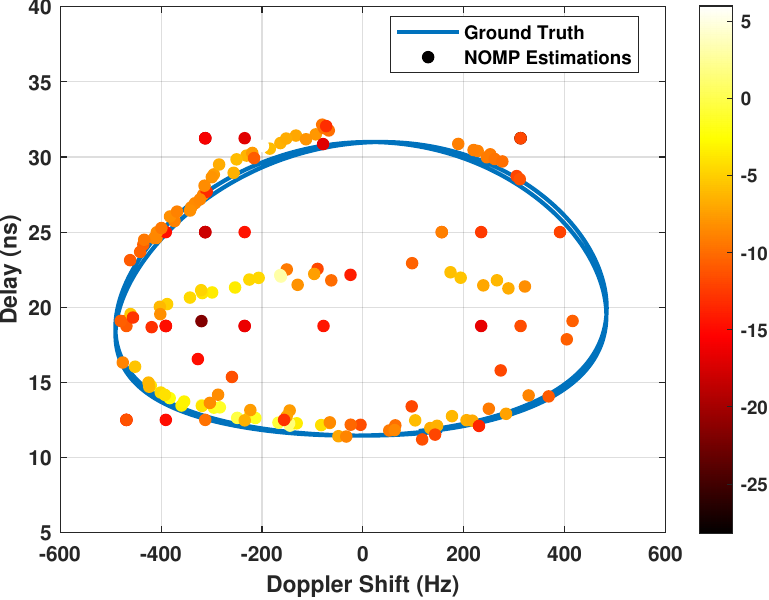}}\\
  \subcaptionbox{Setup 3}{\includegraphics[width=3in,height=\textwidth,keepaspectratio]{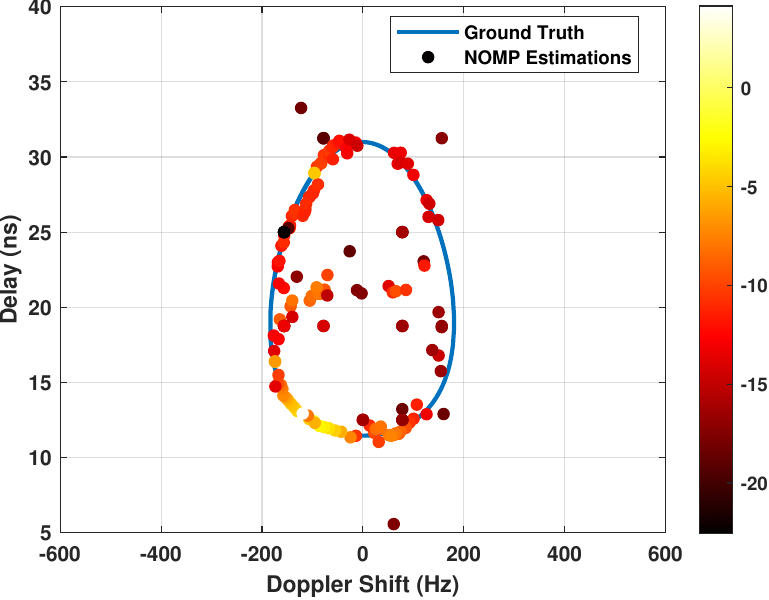}}\quad
  \subcaptionbox{Setup 4}{\includegraphics[width=3in,height=\textwidth,keepaspectratio]{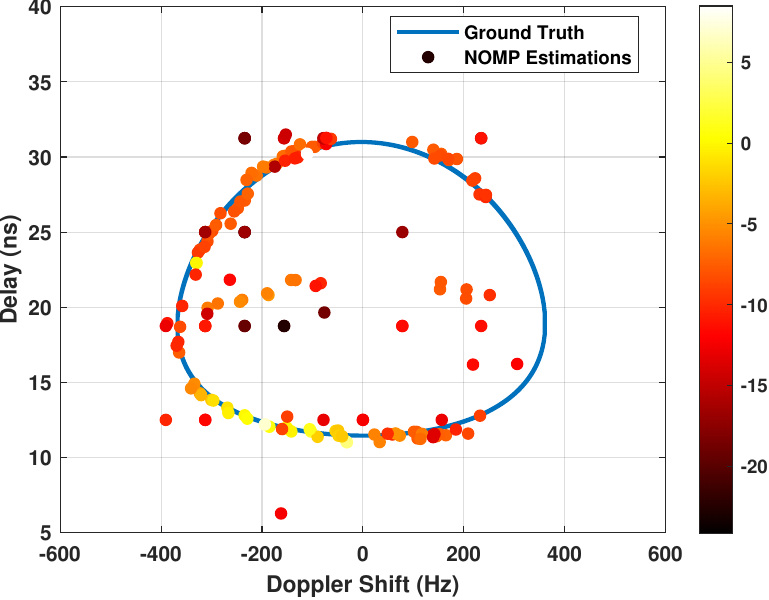}}
  \caption{Comparison of delay and Doppler shift estimations in (a) Setup 1, (b) Setup 2, (c) Setup 3, and (d) Setup 4 with the ground truths for one complete rotation of the target emulator. The color of the estimations represents the magnitude of the moving spheres throughout the complete rotation.}\label{fig:foursetups}
\end{figure*}

\cref{fig:foursetups} presents a comparison of delay and Doppler shift estimation results using the proposed \ac{NOMP}-based algorithm across all four setups, along with the analytically calculated delay and Doppler shift (i.e., ground truths) for one complete rotation of the target emulator.
The \ac{NOMP} utilizes $|\Omega_{\rm{s}}|=\SI[round-precision=2]{1}{\percent}$ of the total time and frequency samples in $\mathbf{H}_c$ for the delay and Doppler shift estimations in all four setups.
Since the measurements are conducted in a semi-anechoic chamber, reflections from the test environment are reduced.
Additionally, reflections from static objects (e.g., motor and antenna arch in VISTA, etc.) are minimized by using the background subtraction algorithm, ensuring that only reflections from the two moving spheres are processed.
Consequently, the sparsity level in the \ac{NOMP} is set to $2$ in the estimation results shown.
Across the four setups, the \ac{NOMP} successfully estimates the delay and Doppler shift of the moving spheres with high accuracy, even when using only $\eta=\SI[round-precision=2]{1}{\percent}$ resources.
This efficiency is evident in the close alignment of the estimations with the ground truths, underscoring the algorithm's performance in different scenarios involving different target sizes and angular speeds.
The missing estimation points in all setups highlight a critical observation regarding the use of directional horn antennas in the measurement setup.
As the spheres rotate, they periodically leave the antennas' field of view, causing the reflected signal power to drop below detectable levels.
This effect is more pronounced in setups with higher angular speeds, especially in setup 2, where the spheres move at \SI{80}{\mathrm{rpm}} through the antenna's field of view, reducing the likelihood of continuous detection.

\begin{figure*}
\centering
  \subcaptionbox{Setup 2}{\includegraphics[width=3in,height=\textwidth,keepaspectratio]{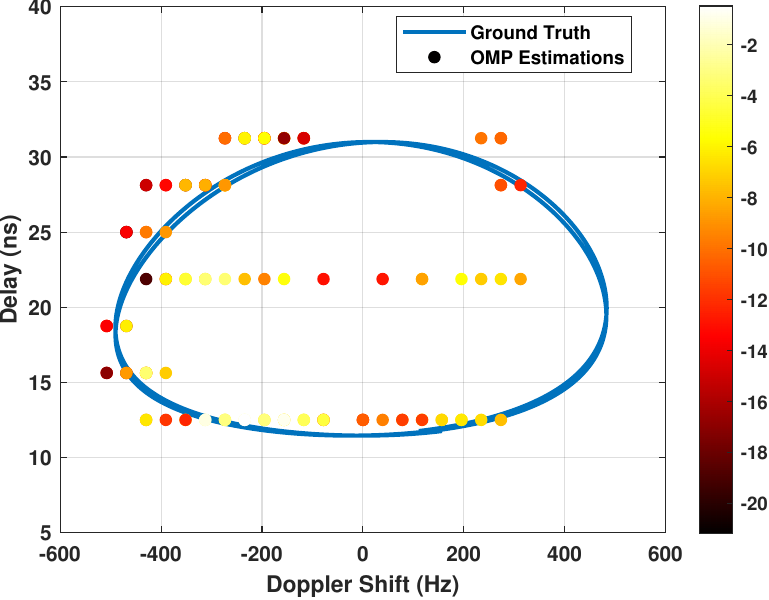}}\quad
  \subcaptionbox{Setup 4}{\includegraphics[width=3in,height=\textwidth,keepaspectratio]{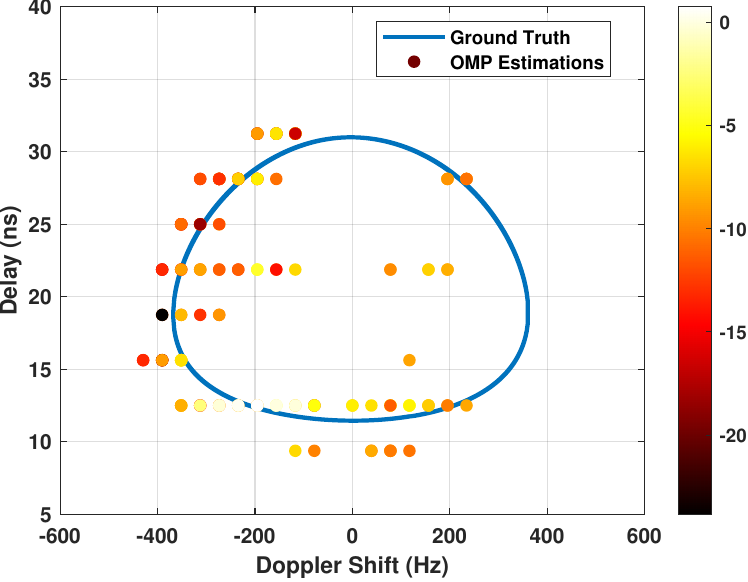}}
  \caption{Delay and Doppler shift estimations results in (a) Setup 2 and (b) Setup 4, using \ac{OMP} with oversampled grid ($\gamma=4$). The color of the estimations represents the magnitude of the moving spheres throughout the complete rotation.}\label{fig:twosetups_omp}
\end{figure*}

In addition to the results obtained using the \ac{NOMP} algorithm, delay and Doppler shift estimations are also performed using the \ac{OMP} algorithm, with an oversampled grid ($\gamma=4$), for setups 2 and 4 as shown in~\cref{fig:twosetups_omp}.
Both setups are evaluated under the same conditions as in the \ac{NOMP} case to enable a direct comparison between the two algorithms.
The results illustrate that the \ac{OMP} algorithm, while capable of performing estimations, suffers from the inherent limitation of on-grid estimations.
Due to the discretization of the parameter space, \ac{OMP} relies on a fixed grid, and as a result, its estimations tend to snap to the closest grid point, leading to inaccurate target detection.
This is due to the basis mismatch problem not handled by the \ac{OMP} algorithm.
In contrast, the \ac{NOMP} overcomes this problem by introducing Newton refinements steps resulting in superior detection accuracy as evident from the much closer alignment between the \ac{NOMP} estimations and the ground truths.
This comparison underscores the limitations of traditional grid-based approaches like \ac{OMP} in applications where high precision is required, especially in high mobility \ac{ISAC} scenarios. 

\begin{figure}[htbp]
\centering
\begin{subfigure}[b]{0.4\textwidth}
   \includegraphics[width=\textwidth]{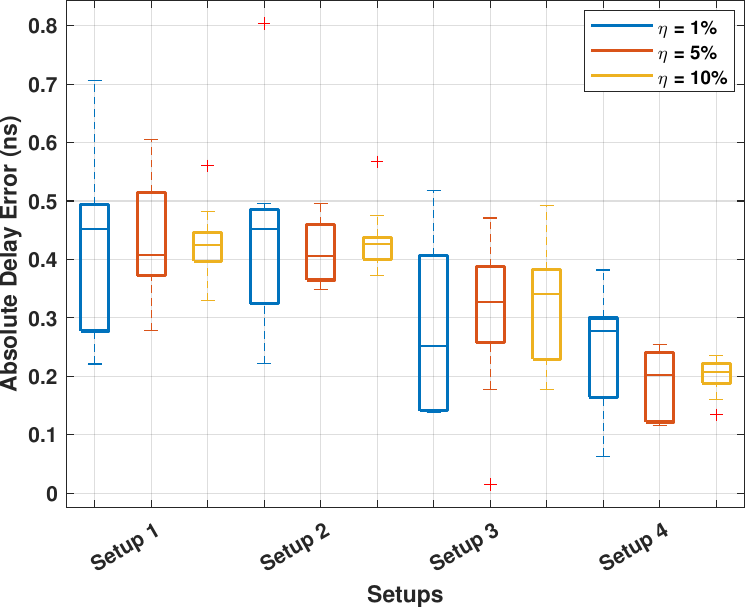}
   \caption{Box plots of absolute delay error distribution across $4$ setups.}
   \label{fig:boxdelay} 
\end{subfigure}
\begin{subfigure}[b]{0.4\textwidth}
   \includegraphics[width=\textwidth]{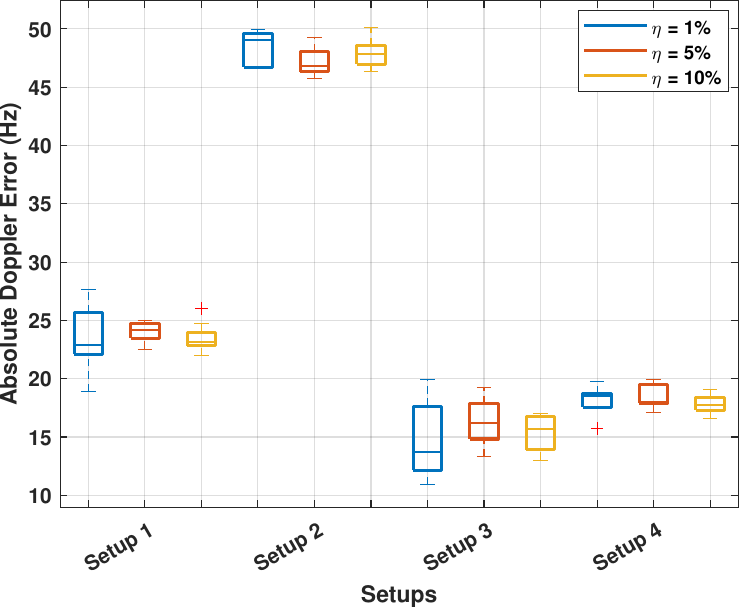}
   \caption{Box plots of absolute Doppler error distribution across $4$ setups.}
   \label{fig:boxdoppler}
\end{subfigure}
\caption{(a) Setups 3 and 4, which feature larger spheres with larger \ac{RCS}, exhibit lower median and overall error compared to setups 1 and 2 with smaller spheres. (b) Setup 2 with the fastest moving spheres shows the largest Doppler estimation error compared to other setups.}\label{fig:boxplot}
\end{figure}

To evaluate the accuracy of the estimation results, the error distributions of the estimated delay and Doppler shift across the four setups are plotted using box plots as shown in~\cref{fig:boxplot}.
To further analyze the impact of resource utilization on estimation accuracy, $50$ simulations are performed for each setup using $\eta=\SI[round-precision=2]{1}{\percent}$, $\eta=\SI[round-precision=2]{5}{\percent}$, and $\eta=\SI{10}{\percent}$ of the total time-frequency resources.
For each simulation, the percentage of resources is randomly selected from the $\mathbf{H}_c$ matrix, ensuring a variety of time-frequency combinations are tested.
\cref{fig:boxdelay} presents the box plot of absolute delay errors, while \cref{fig:boxdoppler} shows the absolute Doppler shift errors.
The results indicate that the error in delay estimation is generally lower in setups 3 and 4, which involve spheres with a larger diameter (\SI{12}{\centi\meter}), compared to setups 1 and 2, which feature spheres with a smaller diameter (\SI[round-precision=2]{6}{\centi\meter}).
This suggests that larger spheres having larger \ac{RCS} provide more accurate delay estimations.
A larger \ac{RCS} leads to stronger reflected signals, increasing the SNR at the receiver.
Higher SNR improves the ability of the \ac{NOMP} to distinguish between the signal and noise, leading to a more accurate estimation of delay.
In contrast, smaller spheres with lower \ac{RCS} produce weaker reflections, resulting in lower SNR, and hence, more estimation errors.
Additionally, across all four setups, the delay estimation error consistently decreases as resource utilization increases, demonstrating that greater resource allocation enhances estimation accuracy.
In terms of Doppler shift estimation, the error is most prominent in setup 2, where the spheres rotate at \SI{80}{\mathrm{rpm}}.
The higher angular speed in this setup introduces more variability in the Doppler shift, leading to greater estimation errors compared to the other setups.
A similar trend is observed in Doppler shift estimation errors with respect to resource utilization, where increasing the percentage of utilized resources reduces the error.
These findings highlight the importance of adequate resource allocation in improving the accuracy of delay and Doppler shift estimation, especially under challenging conditions such as higher angular velocities or smaller target sizes.
\acresetall
\section{Conclusion}\label{conclusion}
In this paper, we proposed a high-resolution off-grid target detection algorithm based on the Newtonized Orthogonal Matching Pursuit (NOMP).
The proposed algorithm is suitable for a multiuser Integrated Sensing and Communication (ISAC) system with arbitrary orthogonal frequency division multiplexing (OFDM) resource grid structure.
The proposed algorithm has three main stages.
In the first stage, the algorithm provides an initial coarse estimate of the target' delay and Doppler shift parameters based on discretized grid points.
In the second stage, this rough on-grid estimate is refined in the surrounding two-dimensional continuous delay-Doppler space using the Newton method.
Finally, in the third stage, a global joint feedback optimization is applied to all estimated targets.
This joint feedback optimization allows inaccurately identified targets with small spacing to be updated and potentially separated more accurately, improving the delay-Doppler resolution. The proposed algorithm offers not only effective and accurate gridless target detection thanks to Newton refinements but also high computational efficiency.
Additionally, the algorithm does not require prior knowledge of the number of targets for termination, which makes it suitable for realistic scenarios.
Besides, the proposed algorithm outperforms the existing typical methods in weak target recovery under strong target components, detection of closely spaced off-grid targets, estimation accuracy, convergence rate, and execution time.
Furthermore, the feasibility of the algorithm is validated using bi-static radar measurement data with dynamic radar targets in a metal-shielded
semi-anechoic chamber.
The measurement results showed that the proposed NOMP-based algorithm can accurately detect targets with only \SI[round-precision=2]{1}{\percent} of the total OFDM resources, significantly reducing the OFDM resource occupancy for radar sensing in ISAC system.

In future work, we can employ the proposed algorithm in resource allocation strategies in ISAC systems.
Given that the algorithm efficiently handles sparse and unstructured OFDM grids, it can be utilized to optimize resource allocation in multiuser ISAC-capable V2X sidelink systems.
By incorporating the algorithm into advanced resource management techniques, the resource occupancy for high-resolution radar sensing can be minimized in dense ISAC systems.
Additionally, extending the algorithm to account for multi-dimensional parameters beyond range and velocity, such as angle-of-arrival, can enable its application in more complex vehicular networks and urban scenarios.
\appendix
In \eqref{blk_Hess_Jacob}, the block Hessian matrix $\mathbf{\ddot S}(\mathcal{C}^\prime)$ and the block Jacobian matrix $\mathbf{\dot S}(\mathcal{C}^\prime)$ are defined.
The entry ${{\mathbf{\dot S}}_{{\ell}}}$ in $\mathbf{\dot S}(\mathcal{C}^\prime)$ is the Jacobian matrix of the ${\ell}^\text{th}$ target and is defined as
\begin{equation}
    \mathbf{\dot S}_{\ell} = \left[ {\begin{array}{c} {\frac{{\partial {\mathbf{S}}(\mathcal{C}^\prime)}}{{\partial \hat\tau_{\ell}^\prime }}} \\ {\frac{{\partial {\mathbf{S}}(\mathcal{C}^\prime)}}{{\partial \hat\alpha_{\ell}^\prime}}} \end{array}} \right],
\end{equation}
where the first order partial derivatives of $\mathbf{S}(\mathcal{C}^\prime)$ w.r.t $\hat\tau_{\ell}^\prime$ and $\hat\alpha_{\ell}^\prime$ are computed as
\begin{align} 
    & \frac{{\partial {\mathbf{S}(\mathcal{C}^\prime)}}}{{\partial \hat\tau_{\ell}^\prime }} = 2\Re \left\{ {{{\left( {{{\mathbf{h}}_r} - {\mathbf{A}}_{\mathcal{C}^\prime}\mathbf{b}_{\mathcal{C}^\prime}} \right)}^{\mathrm{H}}}\frac{{\partial {\mathbf{a}}(\hat\tau_{\ell}^\prime,\hat\alpha_{\ell}^\prime )}}{{\partial \hat\tau_{\ell}^\prime }}\hat\beta_{\ell}^\prime} \right\},\\
    & \frac{{\partial {\mathbf{S}(\mathcal{C}^\prime)}}}{{\partial \hat\alpha_{\ell}^\prime}} = 2\Re \left\{ {{{\left( {{{\mathbf{h}}_r} - {\mathbf{A}}_{\mathcal{C}^\prime}\mathbf{b}_{\mathcal{C}^\prime}}\right)}^{\mathrm{H}}}\frac{{\partial {\mathbf{a}}(\hat\tau_{\ell}^\prime,\hat\alpha_{\ell}^\prime )}}{{\partial \hat\alpha_{\ell}^\prime }}\hat\beta_{\ell}^\prime} \right\},
\end{align}
whereas the entry ${{\mathbf{\ddot S}}_{{\ell}k}}$ in $\mathbf{\ddot S}(\mathcal{C}^\prime)$ is the Hessian matrix, and is defined for two cases: Case 1 when ${\ell}=k$ and Case 2 when ${\ell}\neq k$.
\subsection{Case 1}
When ${\ell}=k$, we have ${{\mathbf{\ddot S}}_{{\ell}{\ell}}}$ as the ${\ell}^\text{th}$ diagonal entry in $\mathbf{\ddot S}(\mathcal{C}^\prime)$, and is defined as
\begin{equation}
    \mathbf{\ddot{S}}_{{\ell}{\ell}} = \left[ {\begin{array}{cc} 
    \frac{\partial^2 \mathbf{S}(\mathcal{C}^\prime)}{\partial \hat{\tau}_{\ell}^{\prime 2}} & 
    \frac{\partial^2 \mathbf{S}(\mathcal{C}^\prime)}{\partial \hat{\tau}_{\ell}^\prime \partial \hat{\alpha}_{\ell}^\prime} \\ 
    \frac{\partial^2 \mathbf{S}(\mathcal{C}^\prime)}{\partial \hat{\alpha}_{\ell}^\prime \partial \hat{\tau}_{\ell}^\prime} & 
    \frac{\partial^2 \mathbf{S}(\mathcal{C}^\prime)}{\partial \hat{\alpha}_{\ell}^{\prime 2}} 
    \end{array}} \right],
\end{equation}
where the second-order derivatives are computed as
\begin{equation}
    \begin{aligned}
        \frac{{\partial^2 \mathbf{S}(\mathcal{C}^\prime)}}{{\partial \hat\tau_{\ell}^{\prime 2}}} & = 2\Re \left\{ \left( \mathbf{h}_r - \mathbf{A}_{\mathcal{C}^\prime}\mathbf{b}_{\mathcal{C}^\prime} \right)^{\mathrm{H}} \frac{{\partial^2 \mathbf{a}(\hat\tau_{\ell}^\prime,\hat\alpha_{\ell}^\prime)}}{{\partial \hat\tau_{\ell}^{\prime 2}}} \hat\beta_{\ell}^\prime\right.\\ 
        & \quad\left. - |\hat\beta_{\ell}^\prime|^2 \frac{{\partial \mathbf{a}^{\mathrm{H}}(\hat\tau_{\ell}^\prime,\hat\alpha_{\ell}^\prime)}}{{\partial \hat\tau_{\ell}^\prime}} \frac{{\partial \mathbf{a}(\hat\tau_{\ell}^\prime,\hat\alpha_{\ell}^\prime)}}{{\partial \hat\tau_{\ell}^\prime}}\right\}, 
    \end{aligned}
\end{equation}
\begin{equation}
    \begin{aligned}
        \frac{{\partial^2 \mathbf{S}(\mathcal{C}^\prime)}}{{\partial \hat\alpha_{\ell}^{\prime 2}}} & = 2\Re \left\{ \left( \mathbf{h}_r - \mathbf{A}_{\mathcal{C}^\prime}\mathbf{b}_{\mathcal{C}^\prime} \right)^{\mathrm{H}} \frac{{\partial^2 \mathbf{a}(\hat\tau_{\ell}^\prime,\hat\alpha_{\ell}^\prime)}}{{\partial \hat\alpha_{\ell}^{\prime 2}}} \hat\beta_{\ell}^\prime\right. \\
        & \quad \left. - |\hat\beta_{\ell}^\prime|^2 \frac{{\partial \mathbf{a}^{\mathrm{H}}(\hat\tau_{\ell}^\prime,\hat\alpha_{\ell}^\prime)}}{{\partial \hat\alpha_{\ell}^\prime}} \frac{{\partial \mathbf{a}(\hat\tau_{\ell}^\prime,\hat\alpha_{\ell}^\prime)}}{{\partial \hat\alpha_{\ell}^\prime}} \right\},
    \end{aligned}
\end{equation}
\begin{equation}
    \begin{aligned}
        \frac{\partial^2 \mathbf{S}(\mathcal{C}^\prime)}{\partial \hat\tau_{\ell}^\prime \partial \hat\alpha_{\ell}^\prime} & = \frac{\partial^2 \mathbf{S}(\mathcal{C}^\prime)}{\partial \hat\alpha_{\ell}^\prime \partial \hat\tau_{\ell}^\prime} \\
        & = 2\Re \left\{ \left( \mathbf{h}_r - \mathbf{A}_{\mathcal{C}^\prime}\mathbf{b}_{\mathcal{C}^\prime} \right)^{\mathrm{H}} \frac{\partial^2 \mathbf{a}(\hat\tau_{\ell}^\prime,\hat\alpha_{\ell}^\prime)}{\partial \hat\alpha_{\ell}^\prime\partial \hat\tau_{\ell}^\prime} \hat\beta_{\ell}^\prime \right. \\
        & \quad \left. -|\hat\beta_{\ell}^\prime|^2 \frac{\partial \mathbf{a}^{\mathrm{H}}(\hat\tau_{\ell}^\prime,\hat\alpha_{\ell}^\prime)}{\partial \hat\alpha_{\ell}^\prime} \frac{\partial \mathbf{a}(\hat\tau_{\ell}^\prime,\hat\alpha_{\ell}^\prime)}{\partial \hat\tau_{\ell}^\prime} \right\}.
    \end{aligned}
\end{equation}
\subsection{Case 2}
When ${\ell}\neq k$, we have ${{\mathbf{\ddot S}}_{{\ell}k}}$ as the $({\ell},k)-\text{th}$ off-diagonal entry in $\mathbf{\ddot S}(\mathcal{C}^\prime)$, and is defined as
\begin{equation}
    \mathbf{\ddot{S}}_{{\ell}k} = \left[ {\begin{array}{cc} 
    \frac{\partial^2 \mathbf{S}(\mathcal{C}^\prime)}{\partial \hat{\tau}_k^{\prime}\partial \hat{\tau}_{\ell}^{\prime}} & 
    \frac{\partial^2 \mathbf{S}(\mathcal{C}^\prime)}{\partial \hat{\tau}_k^\prime \partial \hat{\alpha}_{\ell}^\prime} \\ 
    \frac{\partial^2 \mathbf{S}(\mathcal{C}^\prime)}{\partial \hat{\alpha}_k^\prime \partial \hat{\tau}_{\ell}^\prime} & 
    \frac{\partial^2 \mathbf{S}(\mathcal{C}^\prime)}{\partial \hat{\alpha}_k^{\prime}\partial \hat{\alpha}_{\ell}^{\prime}} 
    \end{array}} \right],
\end{equation}
where the second-order derivatives are computed as
\begin{equation}
    \begin{aligned}
        \frac{{\partial^2 \mathbf{S}(\mathcal{C}^\prime)}}{{\partial \hat\tau_k^{\prime}}{\partial \hat\tau_{\ell}^{\prime}}} & = -2\Re \left\{\left( \frac{{\partial \mathbf{a}(\hat\tau_k^\prime,\hat\alpha_k^\prime)}}{{\partial \hat\tau_k^\prime}}\hat\beta_k^\prime\right)^{\mathrm{H}} \frac{{\partial \mathbf{a}(\hat\tau_{\ell}^\prime,\hat\alpha_{\ell}^\prime)}}{{\partial \hat\tau_{\ell}^\prime}}\hat\beta_{\ell}^\prime\right\}, 
    \end{aligned}
\end{equation}
\begin{equation}
    \begin{aligned}
        \frac{{\partial^2 \mathbf{S}(\mathcal{C}^\prime)}}{{\partial \hat\alpha_k^{\prime}}{\partial \hat\alpha_{\ell}^{\prime}}} & = -2\Re \left\{\left( \frac{{\partial \mathbf{a}(\hat\tau_k^\prime,\hat\alpha_k^\prime)}}{{\partial \hat\alpha_k^\prime}}\hat\beta_k^\prime\right)^{\mathrm{H}} \frac{{\partial \mathbf{a}(\hat\tau_{\ell}^\prime,\hat\alpha_{\ell}^\prime)}}{{\partial \hat\alpha_{\ell}^\prime}}\hat\beta_{\ell}^\prime\right\}, 
    \end{aligned}
\end{equation}
\begin{equation}
    \begin{aligned}
        \frac{{\partial^2 \mathbf{S}(\mathcal{C}^\prime)}}{{\partial \hat\tau_k^{\prime}}{\partial \hat\alpha_{\ell}^{\prime}}} & = -2\Re \left\{\left( \frac{{\partial \mathbf{a}(\hat\tau_k^\prime,\hat\alpha_k^\prime)}}{{\partial \hat\tau_k^\prime}}\hat\beta_k^\prime\right)^{\mathrm{H}} \frac{{\partial \mathbf{a}(\hat\tau_{\ell}^\prime,\hat\alpha_{\ell}^\prime)}}{{\partial \hat\alpha_{\ell}^\prime}}\hat\beta_{\ell}^\prime\right\}, 
    \end{aligned}
\end{equation}
\begin{equation}
    \begin{aligned}
        \frac{{\partial^2 \mathbf{S}(\mathcal{C}^\prime)}}{{\partial \hat\alpha_k^{\prime}}{\partial \hat\tau_{\ell}^{\prime}}} & = -2\Re \left\{\left( \frac{{\partial \mathbf{a}(\hat\tau_k^\prime,\hat\alpha_k^\prime)}}{{\partial \hat\alpha_k^\prime}}\hat\beta_k^\prime\right)^{\mathrm{H}} \frac{{\partial \mathbf{a}(\hat\tau_{\ell}^\prime,\hat\alpha_{\ell}^\prime)}}{{\partial \hat\tau_{\ell}^\prime}}\hat\beta_{\ell}^\prime\right\}, 
    \end{aligned}
\end{equation}
%
%
% \sloppy
% \printbibliography
% \fussy
% \bibliographystyle{ieeetr}
% \bibliography{libs/references}

%
%
%
\begin{IEEEbiography}
[{\includegraphics[width=1in,height=1.25in,clip,keepaspectratio]{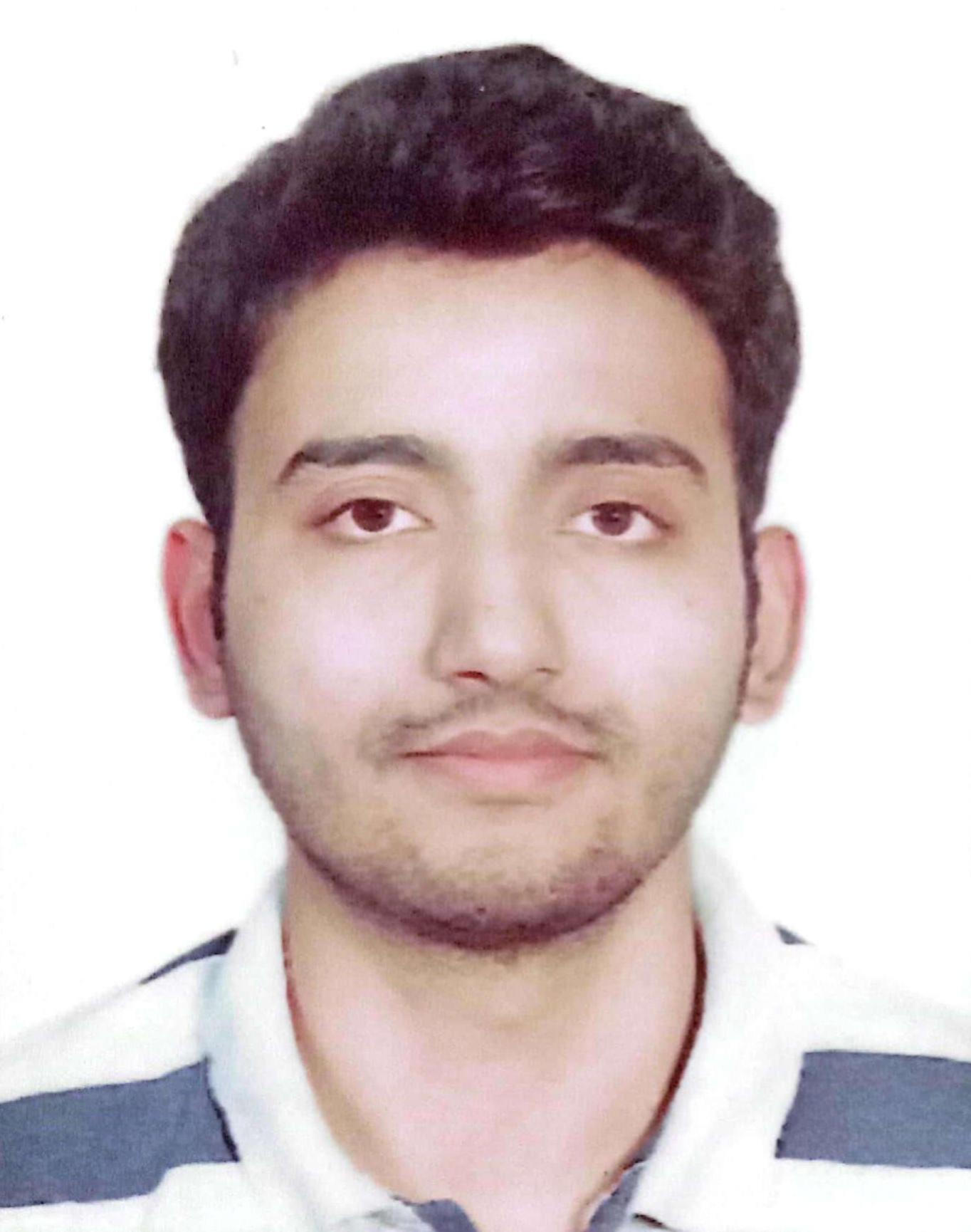}}]{Syed Najaf Haider Shah}
  received his B.Eng. degree in Electrical Engineering from the National University of Sciences and Technology (NUST), Islamabad, Pakistan in 2017 and his Masters degree in Electrical Engineering with majors in digital and wireless communication systems from Information Technology University (ITU), Lahore, Pakistan in 2020.
  Currently, he is pursuing his PhD from Technische Universität Ilmenau (TU Ilmenau), Germany.
  He is associated with the Electronic Measurements and Signal Processing group at TU Ilmenau.
  His research interests include waveform design for joint communication and radar sensing, optimal resource allocation algorithms design in vehicular networks, compressive sensing, and parameter estimation.
\end{IEEEbiography}
\begin{IEEEbiography}[{\includegraphics[width=1in,height=1.25in,clip,keepaspectratio]{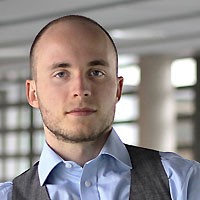}}]{Dr.-Ing. S. Semper}
  studied mathematics at Technische Universität Ilmenau, (TU Ilmenau), Ilmenau, Germany. He received the Master of Science degree in 2015. Since 2015, he has been a Research Assistant with the Electronic Measurements and Signal Processing Group, which is a joint research activity between the Fraunhofer Institute for Integrated Circuits IIS and TU Ilmenau, Ilmenau,
  In 2022 he finished his doctoral studies and received the doctoral degree with honors in electrical engineering. Since then, he has been a post doctoral student in the Electronic Measurements and Signal Processing Group.
  His research interest consist of compressive sensing, parameter estimation, optimization, numerical methods and algorithm design.
\end{IEEEbiography} 

\begin{IEEEbiography}
[{\includegraphics[width=1in,height=1.25in,clip,keepaspectratio]{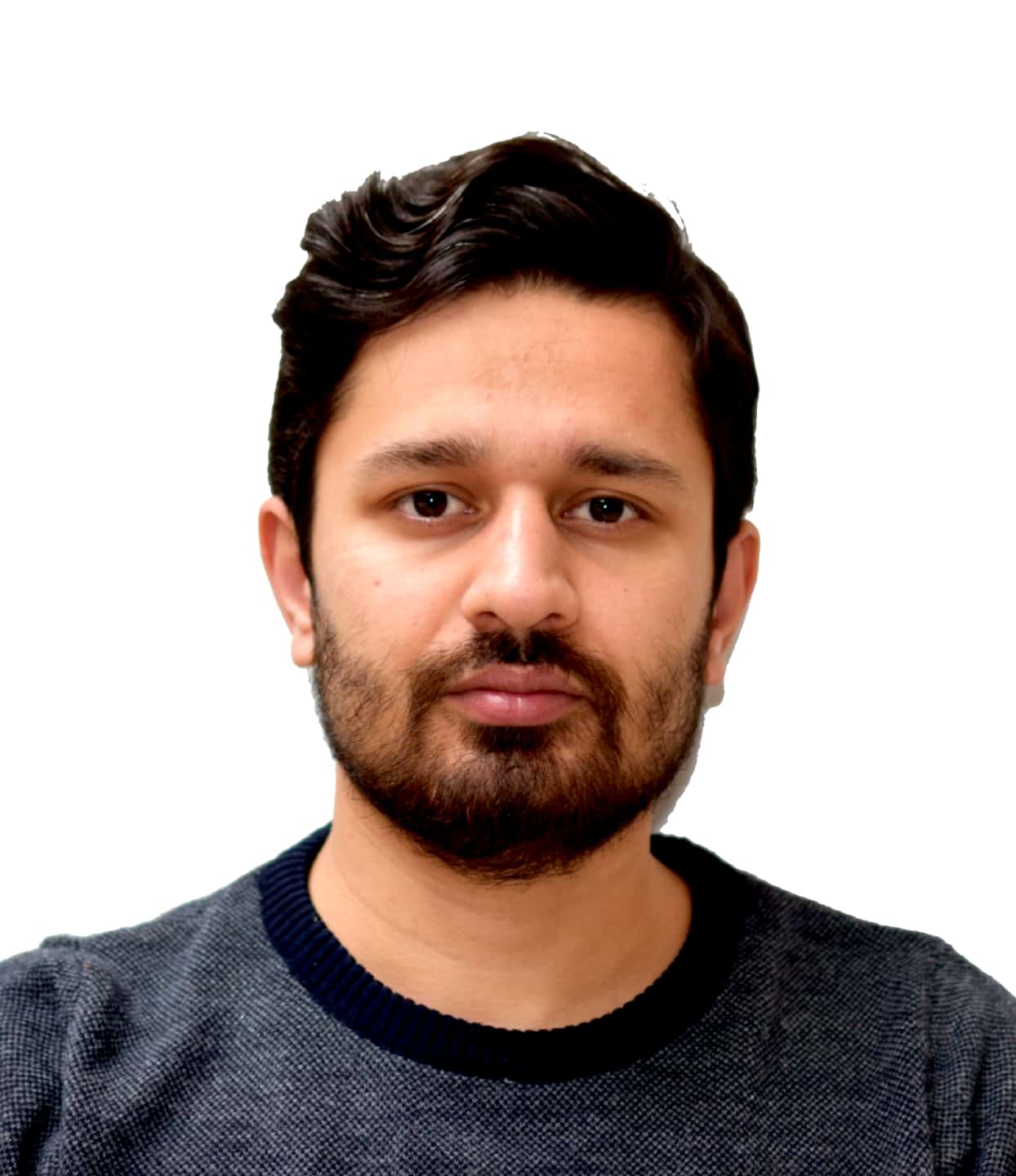}}]{Aamir Ullah Khan}
  received his BS degree in Electrical (Telecommunication) Engineering from COMSATS Institute of Information Technology (now known as COMSATS University (CUI)), Islamabad, Pakistan in 2015, and his Master's degree in Electronics and Communications Engineering from Kocaeli University, Kocaeli, Türkiye.
  He is currently pursuing his PhD degree at Technische Universität Ilmenau (TU Ilmenau), Ilmenau, Germany.
  He is associated with the Electronic Measurements and Signal Processing group at TU Ilmenau.
  His research interests include propagation modeling for Integrated Communication and Sensing (ICAS) enabled Vehicle-to-Vehicle (V2V) systems, parameter estimation, and target detection in a clutter-rich environment. 
\end{IEEEbiography}

\begin{IEEEbiography}
[{\includegraphics[width=1in,height=1.25in,clip,keepaspectratio]{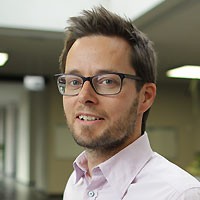}}]{Christian Schneider}
received his Diploma degree in electrical engineering from the Technische Universität Ilmenau, Germany in 2001. 
He is currently a group leader at the Electronic Measurements and Signal Processing department (EMS) at the TU Ilmenau as well as at Fraunhofer IIS.
His research interests include multi-dimensional channel sounding, radio channel characterization and modeling, and its application to space-time signal processing and integrated sensing \& communication (ISAC) questions.
He received a best paper award at the European Wireless Conference in 2013 and the European Conference of Antennas and Propagation in 2017 and 2019.
\end{IEEEbiography}

\begin{IEEEbiography}
[{\includegraphics[width=1in,height=1.25in,clip,keepaspectratio]{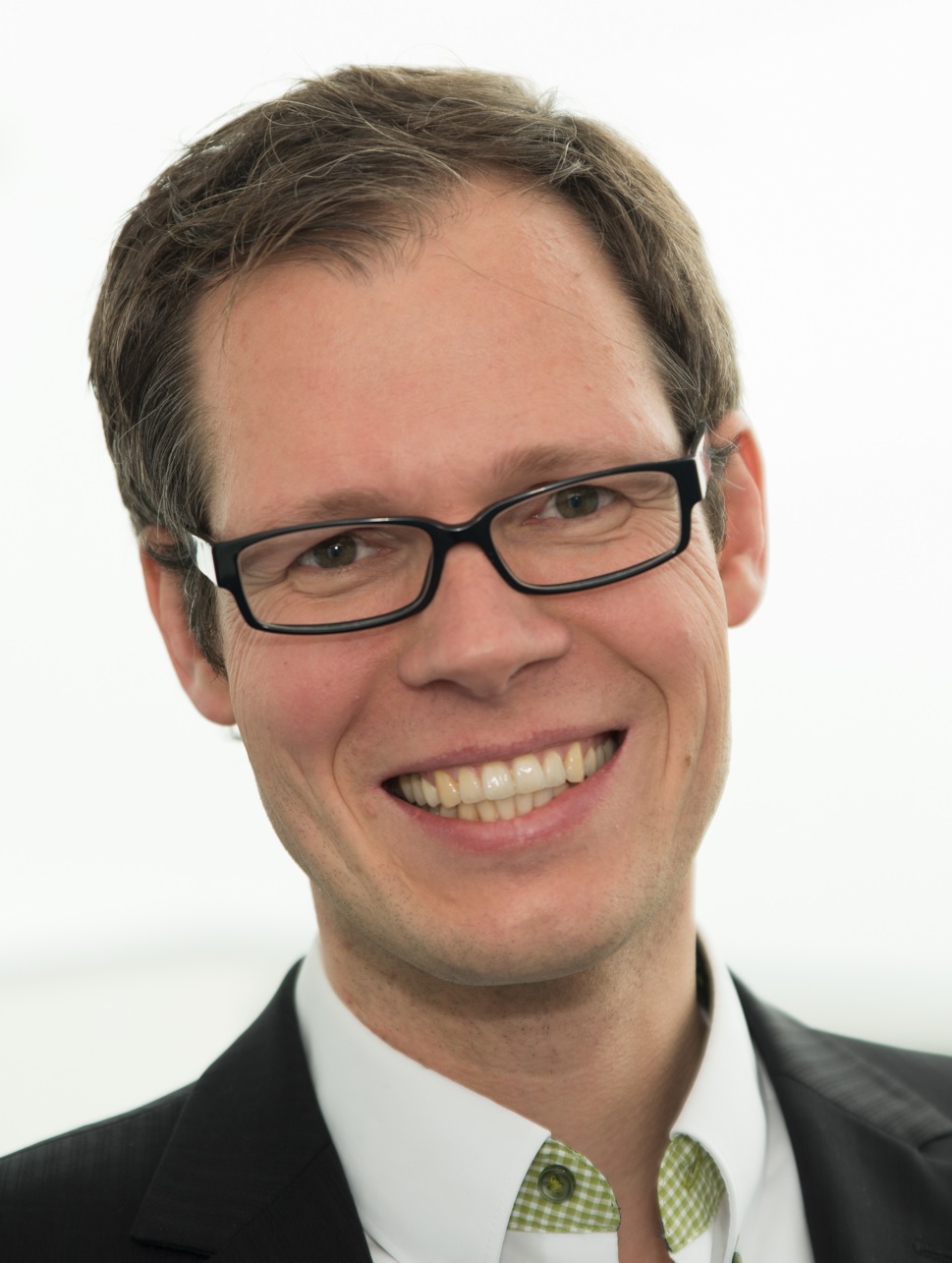}}]{Joerg Robert}
studied electrical engineering and information technology at TU Ilmenau and TU Braunschweig, Germany.
From 2006 to 2012, he conducted research on the topic of digital television at the Institute of Communications Engineering at TU Braunschweig.
Here he was actively involved in the development of DVB-T2, the second generation of digital terrestrial television.
In 2013, he completed his PhD at TU Braunschweig on the topic of \"Terrestrial TV Broadcast using Multi-Antenna Systems\".
In 2012, he joined the LIKE chair at the Friedrich-Alexander-Universität Erlangen-Nürnberg, Germany.
One of his research topics focused on LPWAN (Low Power Wide Area Networks).
Since 2021, he is full professor and head of the Group for Dependable Machine-to-Machine Communication at the Department of Electrical Engineering and Information Technology at Technische Universitaet Ilmenau, Germany.
Joerg Robert is very active in international standardization.
He currently is the secretary of the IEEE 802.15 standardization group, and the vice-chair of the IEEE 802.19.3 Task Group on wireless coexistence.
\end{IEEEbiography}
\end{document}